\documentclass[aps,superscriptaddress,twocolumn,showpacs,secnumarabic,a4paper,floatfix]{revtex4}
\usepackage{amssymb}
\usepackage{color}
\usepackage{amsbsy}
\usepackage{graphicx} 
\usepackage{times,mathptmx}

\newcommand{\nvec}{\mathbf{n}}
\newcommand{\cvec}{\mathbf{c}}

\newcommand{\lvec}{\mathbf{l}}
\newcommand{\mvec}{\mathbf{m}}

\newcommand{\Qvec}{\mathbf{Q}}
\newcommand{\Qij}{Q_{ij}}

\newcommand{\dd}{\mathrm{d}}

\newcommand{\Tr}{Tr}

\begin{document}

\title{Liquid crystals boojum-colloids}

\date{\today}

\author{M. Tasinkevych}
\email{miko@mf.mpg.de}
\affiliation{
Max-Planck-Institut f\"ur Intelligente Systeme, Heisenbergstr. 3,
70569 Stuttgart, Germany,}
\affiliation{Institut f\"ur Theoretische und Angewandte Physik,
Universit\"at Stuttgart, Pfaffenwaldring 57, 70569 Stuttgart,
Germany}

\author{N. M. Silvestre}
\email{nunos@cii.fc.ul.pt}
\affiliation{Centro de F\'\i sica Te\'orica e Computacional and Faculdade 
de Ci\^encias da Universidade de Lisboa, Av. Prof. Gama Pinto 2, 1649-003 Lisboa, 
Portugal}

\author{M. M. Telo da Gama}
\email{margarid@cii.fc.ul.pt}
\affiliation{Centro de F\'\i sica Te\'orica e Computacional and Faculdade 
de Ci\^encias da Universidade de Lisboa, Av. Prof. Gama Pinto 2, 1649-003 Lisboa, 
Portugal}

\begin{abstract} 
 Colloidal particles dispersed in a liquid crystal lead to distortions of the director field. The distortions are responsible for long-range effective colloidal interactions whose asymptotic behaviour is well understood. 
 The short distance behaviour of the interaction, however, is sensitive to the structure and dynamics of the topological defects nucleated near the colloidal particles in the strong anchoring regime. The full non-linear theory is 
required in order to determine the interaction at short separations. Spherical colloidal particles with 
sufficiently strong planar degenerate anchoring nucleate a pair of antipodal surface topological defects,
known as boojums. We use the Landau-de Gennes formalism in order to resolve the mesoscopic structure of the boojum cores and to determine the pairwise colloidal interaction. We compare the results in three ($3D$) and two ($2D$) spatial dimensions. 
The corresponding free energy functionals are minimized numerically using finite elements with adaptive meshes. Boojums are always point-like in $2D$, but  acquire a rather complex structure in $3D$ which depends  on the combination of the anchoring potential, the radius of the colloid, the temperature and the LC elastic anisotropy. We identify three types of defect cores in $3D$ which we call single, double and split core boojums, and investigate the associated structural transitions. The split core structure is favoured by low temperatures, strong anchoring and small twist to splay or bend ratios. For sufficiently strong anchoring potentials characterised by a well-defined uniaxial  minimum, the split  core boojums are the only stable configuration. In the presence of two colloidal particles there are substantial re-arrangements of the defects at short distances, both in $3D$ and $2D$. These re-arrangements lead to qualitative changes in the force-distance profile when compared to the asymptotic quadrupole-quadrupole interaction. In line with the experimental results, the presence of the  defects prevents coalescence of the colloidal particles in $2D$, but not in $3D$ systems.
\end{abstract}


\maketitle

\section{Introduction}

Controlling the self-assembly of colloidal particles is an important aspect of pure and applied
colloid science. In this context, assembling novel structures in liquid crystal matrices has become
a very active field of research as a result of important theoretical and experimental advances 
\cite{Stark.2001,Bohley.2008,Tasinkevych.2010}. Liquid crystals (LCs) are characterized by anisotropic
mechanical and optical properties due to the long-range orientational molecular ordering. Consequently,
in addition to the usual isotropic colloidal interactions, colloids in liquid crystals interact through
strongly anisotropic effective forces, and have been observed to self-assemble into ordered aggregates 
in two ($2D$) \cite{Musevic2006} and three ($3D$) spatial dimensions \cite{ravnik2011}. A distinctive 
feature of these systems is the presence of topological defects, the type of which determines the
symmetry of the far-field director around an isolated colloidal particle, which in turn governs the asymptotic 
behavior of the nematic-induced colloidal interactions \cite{lubensky1998}. The behavior of the interactions 
at short distances is more complex since the defects can move when the colloidal particles are brought close together, 
rendering the description of the self-assembly of liquid crystal colloids a challenging theoretical problem. 

Theoretical and simulation methods have been developed to study LCs on various scales. On the macroscopic
scale, the long-range forces between colloidal particles can be calculated using the Frank-Oseen (FO) 
elastic free energy approach \cite{lubensky1998}. On the mesoscopic scale where non-linear effects become important,
a description based on the Landau-de Gennes (LdG) tensorial order-parameter theory is necessary \cite{ravnik2009}.
This is the case when the interactions between the defects dominate, as revealed by the changes in their equilibrium
structure when the distance between the particles decreases \cite{Tasinkevych.2002}. On the microscopic scale,
presmectic ordering and density variations cannot be ignored and density functional approaches \cite{delasheras2009} 
or computer simulations are  usually employed \cite{andrienko2001}.

The anchoring at the surfaces of colloidal particles can be controlled, for instance, through the adsorption of surfactants or 
deposition of thin organic or inorganic films \cite{kahn1973,jerome1991,Smalyukh.2005a,conradi2009}. 
When the anchoring on a spherical colloid is planar, a tangential nematic director configuration is expected at the colloidal
surface. At the FO level the boundary conditions are met by the creation of a pair of antipodal surface topological defects, called boojums
\cite{Mermin.1977,Mermin.1990}. The boojums are aligned with the far-field director, ensuring that the far-field LC configuration
 is homogeneous as required by the global uniform boundary conditions \cite{Poulin.1998}.

Recently, experimental measurements of the effective forces between two spherical colloids with 
planar anchoring in a $3D$ nematic have been reported \cite{Smalyukh.2005a}. The authors have found that 
the forces at short distances deviate considerably from the asymptotic quadrupole-quadrupole ones, and that
the equilibrium configuration of a pair of particles corresponds to close contact implying
the absence of a repulsion at short distances. By contrast, experimental results obtained for 
inclusions in free-standing smectic $C$ films \cite{Cluzeau.2005,Dolganov.2006} revealed the presence
of a short distance repulsion which keeps the inclusions at a well-defined equilibrium separation larger 
than contact. Again, as in the 3D system of Ref.~\cite{Smalyukh.2005a},
the asymptotic quadrupole-quadrupole approximation for the pair-wise force breaks down at short distances. 

A recent theoretical study  based on the numerical minimization of the $3D$ LdG free energy
functional \cite{Mozaffari.2011} is in line with the experimental results of Ref.~\cite{Smalyukh.2005a} concerning the equilibrium configuration of two colloidal spheres and the absence of colloidal short-range repulsion. 
Theoretical analysis of the $2D$ FO model \cite{Cluzeau.2005,Dolganov.2006}, based on an 
\textit{ad hoc} ansatz for a single particle solution and the superposition approximation, 
have correctly described the configuration of two inclusions in a smectic $C$ film at short distances,
but have failed to provide a consistent mechanism for the short-range repulsion observed experimentally.
A subsequent superposition analysis \cite{Silvestre.2008}, based on the exact single particle solution
\cite{Burylov.1994}, has predicted the existence of two regimes as a function of the anchoring strength. 
For weak anchoring the inclusions coalesce while for strong anchoring, the effective interaction exhibits
a well defined minimum and a strong short-distance repulsion that prevents coalescence \cite{Silvestre.2008}.

Although the $3D$ studies of Ref.~\cite{Mozaffari.2011} are in qualitative agreement with
the experimental results of Ref.~ \cite{Smalyukh.2005a}, the question of the structure of the boojum cores
remains open. The authors of Ref.~\cite{Mozaffari.2011}  could not resolve the structure the defect cores
because the meshes which have been used in the minimization of the LdG free energy functional were too
coarse. In the $2D$ systems, studied in Refs.~\cite{Cluzeau.2005,Dolganov.2006,Silvestre.2008},
the theoretical results are far from satisfactory since the superposition approximation is not valid at short
distances where the interaction between defects is expected. 

Motivated by these open questions, and by the experimental results which reveal a qualitative difference 
between $2D$ and $3D$ systems \cite{Smalyukh.2005,Smalyukh.2005a,Cluzeau.2005,Dolganov.2006},  we 
address in this article the issues of the structure of the boojum cores and the effective interactions
 between circular and spherical colloids in $2D$ 
and $3D$, respectively. We concentrate on the mesoscale and present the results of systematic numerical analysis of the LdG 
free energy functionals in $2D$ and $3D$ for colloidal particles characterized by a tangential (degenerate in $3D$) anchoring.
We use finite element methods (FEM) with adaptive meshes in order to minimize the LdG
functionals and obtain pair-wise effective interactions  at all distances.  We find that the 
defects are point like in $2D$ but acquire a rather complex structure in $3D$. The structure depends on the combination 
of the anchoring potential, the particle radius, the temperature, and the LC elastic anisotropy. More specifically,
we identify defect core transitions between, as we call them, single-core, double-core, and split-core structures.
The split-core structure is favoured  by increasing the anchoring strength or the particle radius and by decreasing 
the temperature. The LC elastic anisotropy responsible for a preferred planar anchoring at the nematic-isotropic interface
(twist elastic constant smaller than bend and splay) \cite{deGennes.1971} also favours the split-core structure. 
Finally, the split-core structure is favored by surface potentials 
characterized by a well-defined uniaxial minimum \cite{Fournier.2005}. In $2D$ the boojums have a single-core
point-like structure for arbitrarily strong surface potentials. 

The paper is organized as follows: In Sec. \ref{3D} we discuss the $3D$ system. First we review the experimental 
results and  define the LdG functional. Then we discuss the results for a single spherical
colloid, focusing on the structures of the boojum core as a function of the anchoring strength, the colloid size and 
the temperature. We consider two types of LCs differing by their anisotropy. Finally we consider the effective interaction 
between two colloidal particles and compare the results with the experimental measurements \cite{Smalyukh.2005,Smalyukh.2005a}.  
In Sec. \ref{2D} we turn our attention to $2D$ systems and calculate the interaction between colloids, where a short-range repulsion is found for strong anchoring. A comparison with the experimental results of 
\cite{Cluzeau.2005,Dolganov.2006} is carried out. In Sec. \ref{conclusions} we discuss and compare 
our results in $2D$ and $3D$. For completeness, the numerical techniques used throughout this article are described in Appendix \ref{appendix}.


\section{Three-dimensional systems}
\label{3D}

In the linear regime the effective force between two colloidal 
particles can be computed within the one-elastic-constant approximation
by using the electrostatic analogy \cite{lubensky1998}.
For the case of spherical colloids with degenerate tangential anchoring
the leading asymptotic term in the multipole ansatz for the director field
is given by a quadrupole term $\propto r^{-5}$ \cite{pergamen2011}, where $r$ is the distance
to the center of the colloidal particle. The superposition approximation 
then leads to the quadrupolar effective interaction between two particles
\cite{Stark.2001}. 

 Smalyukh et al. \cite{Smalyukh.2005a} have studied the
anisotropic interactions between two colloids with tangential anchoring 
by using laser tweezers. At (relatively) large distances $d$ the radial dependence
of the measured force was found to comply with the expected $d^{-5}$ quadrupolar behaviour.
However, the angular dependence of the force disagreed qualitatively with the quadrupolar one
at all distances, indicating the importance of the non-linear effects 
and a breakdown of the superposition approximation. 

In what follows, we shall use the LdG theory
in order to calculate effective interactions between two colloids with tangential anchoring. 
We shall consider two types of LCs differing by their elastic anisotropies.
Strong emphasis is placed on short distance behaviour and strong anchoring limit, 
where the non-linear effects are expected to dominate.
The problem is challenging from the theoretical point of view because the 
defect structure and dynamics are complex, both in the single-particle case
 as well as for interacting particles, as it was predicted a decade ago 
\cite{Tasinkevych.2002} in $2D$ and described recently in $3D$ \cite{Mozaffari.2011}. 
In the following we shall show that the colloidal interaction at short distances results, not
only from the re-arrangement of the defect positions, but also from  structural changes of 
the defect cores. 

Ultimately, we aim at describing quantitatively the experimental results of Smalyukh et al. \cite{Smalyukh.2005a}, 
for the angular dependence of the elastic force at moderate and short inter-particle  distances. 


\subsection{Landau-de Gennes free energy functional}
\label{LdGmodel}
Within the Landau-de Gennes (LdG) theory \cite{Gennes.1993} nematic liquid crystals are
characterised by a traceless symmetric order-parameter tensor with components $\Qij$, 
which can be written as 
\begin{equation}
\Qij = \frac 3 2 Q\left(n_in_j - \frac 1 3 \delta_{ij}\right) + \frac 1 2 B \left(l_il_j - m_i m_j\right) , 
\label{order_parameter} 
\end{equation}
where $n_i$ are the Cartesian components of the director field $\nvec$, $Q$ is the uniaxial order-parameter,
which measures the degree of orientational order along the nematic director, and $B$ is the biaxial order 
parameter, which measures the degree of orientational order along the directions perpendicular to $\nvec$, 
characterized by the eigenvectors $\lvec$ and $\mvec$. The corresponding LdG free energy functional is
\begin{equation}
F_{\mathrm{LdG}} = \int_{\Omega} (f_{b} +
f_{el})\,\dd^3x + \int_{\partial\Omega} f_{s}\,\dd s
\label{free_energy}
\end{equation}
with $f_{b}$ and $f_{el}$ the bulk and elastic free energy densities, given by
\begin{eqnarray}
 f_{b} &=& a \Tr \Qvec^2 - b \Tr
\Qvec^3 + c \left(\Tr \Qvec^2\right)^2 \label{bulk},\\ 
f_{el} &=&
\frac{L_1}{2}\partial_k \Qij\partial_k \Qij + \frac{L_2}{2} \partial_j \Qij
\partial_k Q_{ik}, 
\label{elastic}
\end{eqnarray}
where $a$ depends linearly on the temperature $T$ and is usualy written as $a = a_0 (T - T^*)$, with $a_0$ a 
material dependent constant and $T^*$ the supercooling temperature of the isotropic phase. $b$ and $c$ are 
positive (material dependent) constants, and $L_1$ and $L_2$ are phenomenological parameters which can be related to 
the Frank-Oseen (FO) elastic constants. The first integral in Eq.~(\ref{free_energy}) is taken over the $3D$ domain, $\Omega$,
occupied by nematic, while the second integral is over the surfaces $\partial\Omega$ (in our case the surfaces
of the colloidal particles) and accounts for non-rigid anchoring conditions.

Depending on the preferred orientation of the director (anchoring direction) with respect to the
surface normal $\nu$, the surface anchoring is i) homeotropic, when the anchoring direction is parallel
to $\nu$, ii) planar, when the anchoring direction is orthogonal to $\nu$, and iii) tilted, when the 
anchoring direction and $\nu$ form an angle smaller than $\pi/2$ (and larger than zero).
The last two cases can be classified futher as monostable, multistable or degenerate
(in some articles the term ``random'' is used instead of ``degenerate''),
depending on whether the surface imposes one, a finite or an infinite number of equivalent 
anchoring directions \cite{jerome1991}, respectively.
The simplest quadratic surface free energy, $f_{s}$, favouring monostable nematic ordering ${\Qvec^s}$, i.e.,
with a well-defined director, scalar and biaxial order-parameters was proposed by Nobili and Durand \cite{Nobili.1992}:
\begin{equation}
f_s=W\left(\Qij-\Qij^s\right)^2 
\label{Nobili}
\end{equation}
When $W >0$, the surface free energy $f_s$ has a unique minimum at $\Qvec = \Qvec^s$. In general this is not 
necessary, the requirement being that the total free energy is bounded from below.
In this paper we consider only planar degenerate anchoring 
described by a family of covariant surface potentials 
originally proposed by Fournier and Galatola \cite{Fournier.2005}:
\begin{equation}
f_s=W_1\left(\tilde{Q}_{ij}-\tilde{Q}_{ij}^\perp\right)^2 + W_2\left(\tilde{Q}_{ij}^2 -\Bigl(\frac{3Q_b}{2}\Bigr)^2 \right)^2 
\label{Fournier}
\end{equation} 
where $\tilde{Q}_{ij}=\Qij+Q_b\frac{\delta_{ij}}{2}$, 
$\tilde{Q}_{ij}^\perp=\left(\delta_{ik}-\nu_i\nu_k\right)\tilde{Q}_{ij}\left(\delta_{lj}-\nu_l\nu_j\right)$,
$W_1$ is the anchoring strength favouring tangential orientation of the 
director ${\mathbf n}$, and $W_2>0$ guarantees the existence of a minimum for a scalar order-parameter at the
surface equal to its bulk value, $Q_b$. At a flat surface the nematic is uniform and uniaxial everywhere. 
In the original formulation \cite{Fournier.2005} the surface scalar order-parameter is allowed to vary and the biaxiality at a flat surface increases with the difference between the surface and bulk scalar order-parameters. The quartic
surface potential given by Eq.~(\ref{Fournier}) may be viewed as the minimal biaxiality potential characterised
by a well-defined degenerate planar minimum. The quadratic surface potential ($W_2=0$) is the 
covariant version of Eq.~(\ref{Nobili}) the minimum of which depends on the coupling to the bulk nematic.   

In the strong anchoring regime, the nematic director is parallel to the spherical colloidal surface everywhere, 
with an orientation determined by the far field director. In Sec.~\ref{single} we will see that the quartic term 
$W_2$ has a profound effect on the structure of the topological defects, by controlling the coupling to the bulk
nematic through the deviation of the surface and bulk scalar order-parameters. In the flat surface limit the biaxiality vanishes in the strong anchoring regime \cite{Fournier.2005}, while on the surface of large spherical colloids 
point-like singularities with charge +1 split into pairs of point-like singularities with charge +1/2 connected 
by disclination lines.

It is convenient to define the dimensionless temperature $\tau=24ac/b^2$. At $\tau<1$ the uniaxial nematic is stable
and the degree of orientational order is given by
\begin{equation}
Q_b=\frac{b}{8c}\left(1+\sqrt{1-\frac{8\tau}{9}}\right)
\end{equation}
The nematic becomes unstable at $\tau>9/8$. At $\tau=1$ both the nematic and the isotropic phases coexist. 
Typical values of the bulk parameters for 5CB are \cite{kralj.1991} $a_0=0.044\times10^6$ J $/$Km$^3$, 
$b=0.816\times10^6$ J$/$m$^3$, and $c=0.45\times10^6$ J$/$m$^3$, $L_1=6\times10^{-12}$J$/$m,
$T^* = 307$ K. The spatial extension of inhomogeneous regions and the cores of topological defects 
is of the order of the bulk correlation length, which is given by $\xi=\left(8c\left(3L_1+2L_2\right)/b^2\right)^{1/2}$ at the 
nematic-isotropic (NI) transition \cite{Chandresakhar.1992}.

We define the elastic constant anisotropy $\eta \equiv L_2/L_1$, and consider two 
cases i) $\eta = 2$ with $\xi\simeq15\mathrm{nm}$ corresponding to 5CB, and $\eta = -1/2$ 
with $\xi\simeq8\mathrm{nm}$. Note that stability arguments require $\eta > -3/2$ \cite{Longa.1987}.
Depending on the LC material, $L_2$ can be positive or negative, and its sign controls the molecular
 orientation at the NI interface, which is planar for $\eta >0$ and homeotropic for $\eta <0$ \cite{deGennes.1971}.  

The LdG elastic constants $L_1$ and $L_2$ may be related to the FO elastic constants \cite{Frank.1958}, 
$K_1=K_3$ and $K_2$, through the uniaxial ansatz $\Qij=\left(3/2\right)Q_b\left(n_i n_j -\delta_{ij}/3 \right)$, 
yielding $K_{1} = K_{3} = 9 Q_b^2(L_1+ L_2/2)/2$ and $K_{2} = 9Q_b^2L_1/2$. In general $K_1$ and $K_3$ are different,
 but in most cases the difference is small and the LdG free energy is deemed adequate.

In the following the LdG free energy Eq.~(\ref{free_energy}) is minimised numerically using FEM
with the adaptive mesh refinement. A detailed description of the numerical procedures is 
given in Appendix \ref{appendix}.


\subsection{Single spherical particle: three types of boojum cores}
\label{single}
When spherical colloids with strong planar anchoring are dispersed in nematic LCs a pair of topological surface defects, 
named {\it boojums}, appears at the antipodes of the particles \cite{Poulin.1998,Mermin.1977,Mermin.1990}.
The understanding of the structure of defects in terms of the order-parameter distribution in the cores 
is beyond mere topological arguments and has been the subject of investigation since the early 1930s \cite{Oseen.1933}. 
For infinitely strong anchoring and within the class of axially symmetric
fields, hedgehog defects exhibit three different structures: the radial hedgehog, a small ring or loop disclination 
\cite{Penzenstadler.1989} and a third structure, a split-core defect, which was found to be metastable \cite{Mkaddem.2000}. 
Phase and bifurcation diagrams indicate that the transition from the hedgehog to the ring structure is first order, 
as predicted in Ref.~\cite{Penzenstadler.1989}.
Although several studies addressed the structure of hedgehog defect cores \cite{Schopohl.1988,Penzenstadler.1989,Mkaddem.2000} 
the structure of boojums has not been fully understood \cite{Kralj.2008}. 

A homotopy classification of surface topological defects is given by Volovik in Ref.~\cite{Volovik.1978}. According 
to Volovik's classification a point defect at the boundary of a nematic LC may be viewed as the combination of a bulk 
hedgehog and a surface boojum and is therefore characterised by two topological charges -- the charge $N$ of the bulk 
hedgehog and the index $m$ of the projection $\mathbf{t = n - \nu (n \cdot \nu) }$ of the director field onto 
the surface with normal $\nu$. The charges $N$ and $m$ are related to a continuous 
topological charge ${\cal A}$ defined as \cite{Volovik.1983,kurik:1988}
 \begin{equation}
{\cal A}=\frac{1}{4\pi}\int_\sigma{ \nvec\left(\frac{\partial\nvec}{\partial\theta}\times\frac{\partial\nvec}{\partial\phi}\right)
d\theta d\phi}
\label{volovik}
\end{equation}
where $\sigma$ is a hemisphere surrounding the defect, and $\theta,\phi$ are arbitrary coordinates on $\sigma$. 
Evaluation of the integral in Eq.~(\ref{volovik}) gives an explicit relation between the continuous charge ${\cal A}$ and the charges $N$ and $m$ \cite{Volovik.1983}
\begin{equation}
 {\cal A} = \frac{m}{2} (\mathbf {n\cdot \nu} -1) + N
\end{equation}
which provides a means of calculating the charge $N$ as the index $m$ can be calculated independently \cite{kurik:1988}. 
If ${\cal A}$ is an integer the defects can detach from the surface and if ${\cal A}=0$ they may vanish altogether. 
By contrast, for non-integer charges ${\cal A}$ the defects are ``topologically" bound to the surface. For instance, 
a single-core boojum, which is discussed below, has ${\cal A}=\frac{1}{2}$ and $N=1$, i.e., the defect cannot vanish nor 
detach itself from the surface.

In this section we investigate the dependence of the boojum core structure on i) the strengths of the anchoring potential,
$W_1,W_2$, ii) the reduced temperature $\tau$, and iii) the colloidal radius $R$. We introduce the dimensionless anchoring strengths $w_i=W_i Q_b^2R/K_2$ ($i=1,2$), and consider LCs with positive (e.g. 5CB) and negative elastic anisotropies. We pay particular attention to the nature and degree of the nematic order within the boojum cores. In agreement with Ref.~\cite{Biscari.1997} we find that the apparent singularities of the director field  are replaced by uniaxial order-parameter distributions with negative scalar order-parameter, corresponding to oblate nematic order, surrounded by  biaxial layers.

We shall show that boojum cores are axially symmetric point-like with index $m=+1$ on small colloids.
We call this structure single core boojum. It is stable at high temperatures and relatively weak anchoring. In Ref.~\cite{Kralj.2008} a similar boojum core structure (named by the authors  ``fingered'' boojum) has been reported at a flat surface. On large colloids, at low temperatures and strong anchoring, the axial symmetry is broken and the boojum $+1$ 
point-like cores split into pairs of $+\frac{1}{2}$ point-like surface defects which are connected by disclination lines. 
A structure without a fully developed disclination line, the double core boojum, is also found. The detailed structure of the boojums as well as the transitions between the different configurations depend in detail on the colloid and the LC parameters as we shall discuss below.

\begin{figure}[t]
\centering
\includegraphics[width=0.5\textwidth]{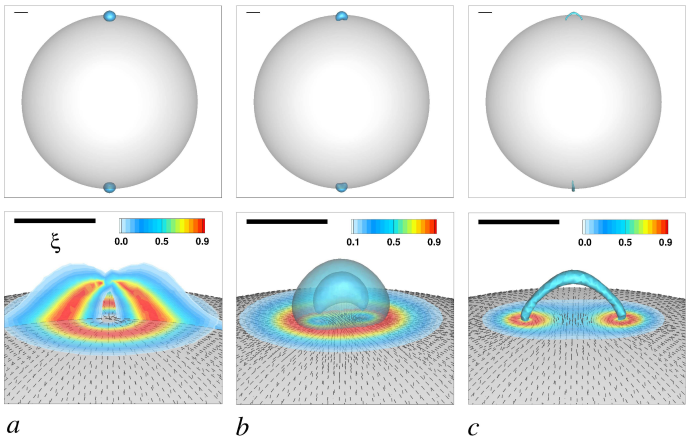}
\caption{Three types of boojum core structures: $a)$ single core, $b)$ double core, and $c)$ split core configurations.
The transitions between different configurations are driven by decreasing temperature $\tau$: a) $\tau\simeq -0.3$, 
$b)$ $\tau\simeq -0.7$, c) $\tau \simeq -3.2$, at fixed anchoring strengths, $w_1 \simeq 37, w_2 =0$, radius of the 
colloidal particle, $R = 0.1 \mu m \simeq 6.6\xi$, and elastic constant anisotropy, $\eta = 2$. The scale bars 
represent the nematic correlation length $\xi$ at the NI transition. In the upper panels and in the lower
panel $c)$ the iso-surface of the scalar order-parameter corresponding to $Q = 0.3 Q_b$ is shown. In the lower 
panel $b)$ the proto-ring of the double core boojum is represented by the iso-surfaces $Q = 0.4 Q_b$ and $Q = 0.2 Q_b$.
The lower panels depict the color-coded biaxiality parameter Eq.~(\ref{biaxiality_param}) in the plane $z=R$. 
The biaxiality parameter in the plane $y=0$ is also shown for the single core boojum, in the lower panel $a)$. 
The short lines on the surface of the colloidal particle represent the eigenvector of $\mathbf Q$ corresponding 
to the largest eigenvalue. This eigenvector corresponds to the nematic director only in the case of prolate nematic
 order. See the main text for further details.}
\label{core_types} 
\end{figure}

We start by describing the three distinct stable configurations of boojum cores, namely single, double and split core boojums. The cores differ both in their surface and bulk structures. Typical core configurations are illustrated in the lower panels of Fig.~\ref{core_types} where the degree of biaxiality defined as 
\begin{equation}
\beta^2 = 1-6\frac{({\mathrm \Tr}{\mathbf Q^3})^2}{({\mathrm \Tr}{\mathbf Q^2})^3}
\label{biaxiality_param} 
\end{equation}
is shown by a color map. $\beta=0$ characterises the uniaxial nematic while $\beta=1$ corresponds to the maximal biaxiality
nematic state. $\beta=1$ is obtained when one (and only one) eigenvalue of $\mathbf Q$ vanishes.
Note that in the isotropic phase all of the eigenvalues of $\mathbf Q$ vanish \cite{Mkaddem.2000}.
Figure \ref{core_types} illustrates the effect of $\tau$ on the structure of boojums for $\eta =2$ and
strong quadratic ($w_2=0$) surface potential. As $\tau$ decreases the boojum core transforms from the single to 
the double core, and then to the split core structure with the accompanying disclination half-ring. 
The core is fairly large close to the NI transition and shrinks as the temperature decreases.

\begin{figure}[t]
\centering
\includegraphics[width=0.48\textwidth]{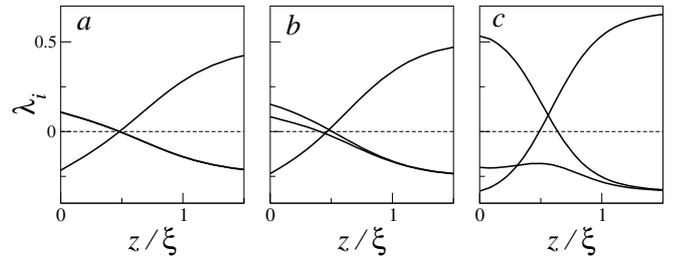}
\caption{Eigenvalues of the tensor order-parameter $Q_{ij}$ on the $z-$axis for the three types of boojum cores: 
$a)$ single core, $b)$ double core, and $c)$ split core configurations, respectively. The model parameters are as in Fig.~\ref{core_types}. The far-field director is oriented along the $z-$axis.} 
\label{core_types_eigenvalues_along_z}
\end{figure}

A description of the boojum core configurations in terms of the eigenvalues of $\mathbf Q$ along the $z$-axis
(the direction of the far-field director) is illustrated in Fig.~\ref{core_types_eigenvalues_along_z}, 
where the eigenvalues $\lambda_i, i=1,2,3$ are plotted as functions of the distance
$z$ to the colloidal surface at fixed $x=y=0$ (the $x,y$ coordinates
of the center of the colloidal particle). 

Single core boojums are uniaxial along the $z-$axis, as revealed by the two degenerate eigenvalues. 
The scalar order-parameter vanishes at $z/\xi\simeq 0.5$ above the colloidal surface, where the three
 eigenvalues vanish and the system is isotropic. Below this point the scalar order-parameter is negative, 
since the principal (non-degenerate) eigenvalue is negative and the nematic order is oblate. 
Above $z/\xi\simeq 0.5$ the scalar order-parameter is positive indicating that the nematic becomes prolate. 

\begin{figure}[t]
\centering
\includegraphics[width=0.48\textwidth]{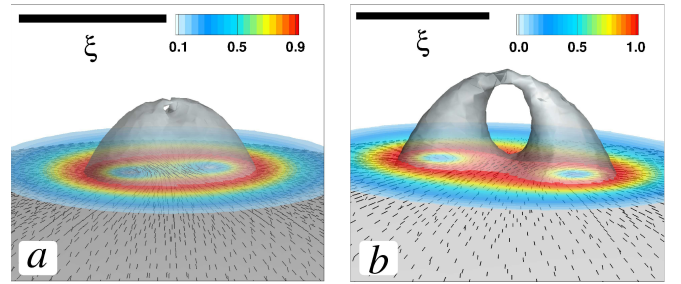}
\caption{Structure of the double core boojum represented by the iso-surface $\det \mathbf Q = 0$, and the color-coded biaxial parameter Eq.~(\ref{biaxiality_param}) in the plane $z=R$.  
$a$: the model parameters are as in Fig.~\ref{core_types}$b$; 
$b$: $\tau\simeq 0.16$, $R\simeq 9.3\xi$.}
 \label{handled_hemisphere}
\end{figure}

By contrast, double core boojums are biaxial along the $z-$axis. The surface of maximal biaxility intersects 
the $z-$axis at three distinct points (see Fig.~\ref{core_types_eigenvalues_along_z}$b$ and Fig.~\ref{handled_hemisphere})
 separating regions of low nematic oblate and prolate order.    

Split core boojums are also biaxial along the $z-$axis. The surface of maximal biaxiality intersects the $z-$axis at two
 distinct points, $z/\xi\simeq 0.49$ and $z/\xi\simeq 0.63$, that delimit the region of oblate nematic order. 
In the outer region, $z/\xi > 1$, two of the eigenvalues are degenerate and the nematic liquid crystal is uniaxial. 
As $z$ decreases the two negative eigenvalues depart from each other, implying non-zero biaxiality. One eigenvalue remains negative. The other two exchange places at a point that can be identified with the center of the boojum core. At this point the nematic is uniaxial with $Q_b<0$. All such points form a $3D$ curve which is surrounded by the surface of maximal biaxiality ($\det \mathbf Q = 0$). This surface encloses the region with oblate uniaxial order ($Q_b<0$). 
 
In all cases, the boojum cores are non-singular and biaxial over extended regions, with linear dimensions of the order 
of the bulk correlation length, $\sim \xi$. 
We proceed to describe in detail the structure of the three types of boojum cores.  

Single core boojums are characterised by an axially symmetric distribution of the order-parameter $\mathbf Q$.
Consequently, $\beta \equiv 0$ on the $z$ axis, where two eigenvalues are degenerate, as shown in 
Fig.\ref{core_types_eigenvalues_along_z}$a$. The iso-surface $\det {\mathbf Q} = 0$ is a distorted hemisphere 
that is biaxial ($\beta=1$) everywhere (See Fig.\ref{core_types}$a$) except at the point where it intersects 
the $z$ axis, $z /\xi\simeq 0.5$, where the fluid is isotropic. At $z = 0$, the location of the putative surface
 point defect, the scalar order-parameter is negative, i.e., the molecules are perpendicular to the director \cite{Gennes.1993} 
(oblate nematic order).

In double core boojums the axial symmetry is broken. As a result the system becomes biaxial along the $z$ axis, 
Figs.~\ref{core_types}$b$ and \ref{core_types_eigenvalues_along_z}$b$. The most prominent feature of this structure
however, is the appearance of a proto-ring resembling the fully developed half-ring  disclination of the split core boojum. 
The proto-ring is formed in a region, of the order of $\xi$, of low nematic order (the scalar order-parameter is 
$\leq 0.5 Q_b$) and is depicted in Fig.~\ref{core_types}$b$ by the iso-surface $Q = 0.2 Q_b$. 
The double core structure shares a number of features with the split core boojum. Its innermost 
core is a half-ring starting and ending on the colloidal surface, with very low biaxiality, $\beta\simeq 0$,
 and negative scalar order-parameter. Further inspection of the eigenvalues of $\mathbf Q$ reveals that the 
iso-surface $\det \mathbf Q = 0$ ($\beta=1$) is a hemisphere with a handle at the top, see Fig.~\ref{handled_hemisphere}.
 This iso-surface intersects the colloidal surface in a single closed loop. In the hole, below the handle, 
the biaxiality is vanishingly small and the scalar order-parameter is positive, i.e., the nematic is prolate uniaxial.

Finally, the split core boojum is characterised by a $\frac{1}{2}$ disclination half-ring connected to the colloidal surface. The inner core of the ring has zero biaxiality and oblate nematic order. This region is surrounded by the iso-surface $\det \mathbf Q = 0$, forming a half-torus see Figs.~\ref{core_types}$c$ and \ref{core_types_eigenvalues_along_z}$c$, which intersects the colloidal surface in two distinct closed loops. The nematic is prolate uniaxial everywhere except in a region of order $\sim \xi$ in the core of the half-ring. 

How does the stability of these structures depend on the colloidal and LC parameters? All three types of
 boojums are topologically equivalent, in the sense that any of them can be transformed smoothly into any
 other. The split and double core boojums result from breaking the axial symmetry of the single core, which 
splits on the colloidal surface. This splitting is complete in the split core boojum in the sense that it is
 accompanied by the splitting of the $\det \mathbf Q = 0$ surface, and only partial in the double core boojum.  
The dependence of the equilibrium (stable) boojum core structure on the strength of the anchoring potential,
 $W_1,W_2$, the reduced temperature $\tau$, and the radius $R$ is complex and 
will be addressed below for LCs with positive and negative elastic anisotropies.

\subsubsection {Strength of the anchoring potential \\}

Usually the anchoring on the surface of the colloidal particle is assumed to be infinitely strong. Here we investigate in detail
the effects of the strength of the quadratic ($w_1$) and quartic ($w_2$) terms of the surface potential, Eq. (\ref{Fournier}).
When the anchoring strengths $w_1$ and $w_2$ are large the split core structure is favoured 
independently of $R,\tau$, and $\eta$. When $w_2=0$ and $w_1$ is large the split (double) core boojum is 
found only for sufficiently large colloids and low temperatures in LCs with $\eta>0$ ($\eta<0$). 
At flat surfaces the quartic
potential induces uniform uniaxial nematic profiles \cite{Fournier.2005}. On spherical colloids, in the strong 
anchoring regime, this potential stabilizes the split core boojums where the biaxial regions are reduced. 
By contrast, in systems with a quadratic surface potential only, $w_2=0$, the minimum is determined by the coupling of
the surface to the bulk nematic \cite{Fournier.2005}. In this case the stability of the boojum cores depends in
detail on  $W_1$, $R$, $\tau$, and $\eta$.  

\begin{figure}[t]
\centering
\includegraphics[width=0.45\textwidth]{figure04_top.eps} \\
\includegraphics[width=0.45\textwidth]{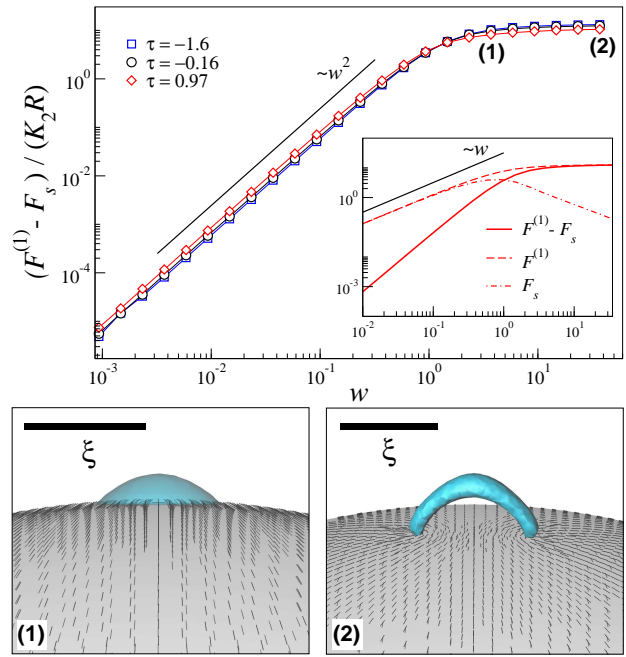}
\caption{Elastic free energy (excess over the uniform nematic free energy $F^{(1)}$ minus the surface free energy $F_s$)
 as a function of the dimensionless quadratic anchoring strength, $w \equiv w_1$, $(w_2 =0)$, at several values of 
reduced temperature $\tau$. The inset shows the elastic (full line) excess (dashed line) and surface (dash-dotted line)
free energies as functions of $w$. The lower panel depicts the configurations of (1) the single core boojum for 
 $w \simeq 3.7$, and (2) the split core boojum for $w \simeq 37$, $\tau\simeq-1.6$. The black lines on the
surface of the colloidal particle represent the eigenvectors corresponding to the largest eigenvalue of ${\mathbf Q}$.
The iso-surfaces corresponding to $Q = 0.3 Q_b$ are also shown. The LC anisotropy $\eta=2$, and the radius of the colloid  
$R/\xi \simeq 6.6$.}
\label{effect_of_W} 
\end{figure}


We start by investigating the quadratic surface potential, $w_2=0$. In Fig.\ref{effect_of_W} we plot the elastic
free energy $F^{(1)}-F_s$, where $F^{(1)}\equiv F_{\mathrm{LdG}} - f_b(Q_b)\Omega$ is the excess (over the uniform uniaxial nematic) free energy and $F_s$ is the surface free energy, as a function of the dimensionless anchoring strength, $w \equiv w_1$, for several values of $\tau$ from just below the NI transition ($\tau\simeq 0.97$) to deep in the bulk nematic
($\tau \simeq -1.6$), at fixed values of $\eta$ and $R$. $\Omega$ is the LC volume.

Two regimes are observed. In the strong anchoring regime, $w \gg 1$, the elastic free energy exhibits a plateau 
that starts with the nucleation of the boojums at opposite poles of the colloidal particle. The inset shows that
in this regime the excess, $F^{(1)}$, and the elastic, $F^{(1)}-F_s$, free energies approach each other asymptotically,
since the surface free energy $F_s$ decreases roughly as $w^{-1}$. In the weak anchoring regime, $w \lesssim 1$, 
the surface free energy increases linearly with $w$ and dominates over the elastic term, which grows as $w^2$. 
This behaviour is in line with the case of the homeotropic anchoring, where the elastic free energy was also found 
to vary as $w^2$ in the weak and to saturate in the strong anchoring regimes \cite{Kuksenok.1996}. 

Figures \ref{effect_of_W}(1) and (2) illustrate how the structure of the boojum cores evolves as
the strength of quadratic surface potential increases, at low temperatures.
As the anchoring strength increases the competition between elastic and anchoring energies leads to
conflicting director orientations at the poles of the colloidal particles where the regions of 
reduced uniaxial and increased biaxial order develop. 
Inspection of the director configurations confirms that the system starts by developing a single core point-like surface defect, see Fig.\ref{effect_of_W}$(1)$. The projection $\mathbf{t}$ of the director field onto 
the surface of the particle 
is characterised by an index $m = +1$. In this regime of small $w$, the surface director field has a finite normal 
component as can be seen in Fig.\ref{effect_of_W}$(1)$.
Deep in the strong anchoring regime the normal component  of the surface director vanishes, see Fig.\ref{effect_of_W}$(2)$,
and the region of reduced uniaxial nematic order opens into a half-ring which is attached to the colloidal surface,
and resembles a $\frac{1}{2}$ disclination line defect. 

\begin{figure}[t]
\centering
\includegraphics[width=0.45\textwidth]{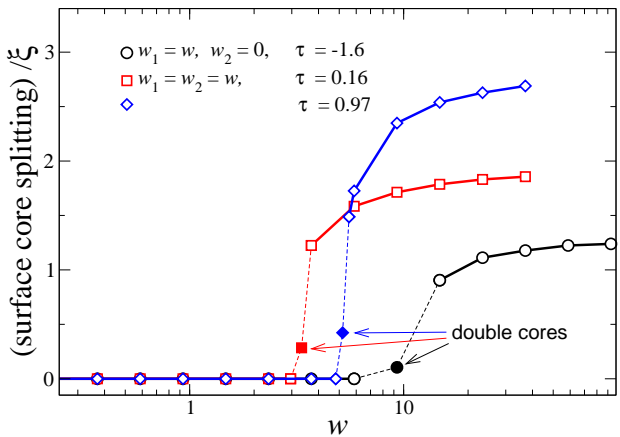}
\caption{Defect core transitions driven by the anchoring strength $w$. Circles correspond
to a quadratic anchoring potential, $w_2 = 0$, while squares and diamonds correspond to a quartic anchoring
potential, $w_1$ and $w_2\neq0$. The surface core splitting is the distance between two $\frac{1}{2}$ point-like
surface defects in the double and split core boojums. Single core boojums are stable at small 
anchoring strengths and high temperatures. Split core structures are stable at large anchoring strengths and 
low temperatures. $\eta = 2$, and the colloid radius $R/\xi \simeq 6.6$. 
Solid symbols correspond to double core boojums that are stable over a narrow
range of anchoring strengths.}
 \label{cores_by_anchoring} 
\end{figure}

For the quadratic surface potential ($w_2 = 0$) boojum core structural transitions driven by the anchoring strength are observed at low temperatures only. The boojum core structure diagram is shown in Fig.~\ref{cores_by_anchoring} by circles. The single core is stable at weak anchoring $w \lesssim 6$, the double core is stable in a narrow range of anchoring strengths, $6\lesssim w \lesssim 15$, while the split core is stable at strong anchoring $w \gtrsim 15$.
At high temperatures ($\tau = 0.16, 0.97$, not shown in Fig.\ref{cores_by_anchoring}) the quadratic
anchoring potential is incapable of stabilizing the double or spilt core boojums. Only the single 
core structure is observed in the range of anchoring strengths $0.00037\lesssim w \lesssim 370$.


We now proceed with the analysis of the quartic potential, Eq. (\ref{Fournier}). We set $\eta=2$,
$R \simeq 6.6\xi$ and assume, for simplicity $w \equiv w_1=w_2$.
We find three types of stable cores, at all temperatures. As the strength $w$ 
of the anchoring potential increases, the sequence of single, double and split core structures is observed. 
Two cases corresponding to $\tau \simeq 0.16$ (squares) and $\tau\simeq 0.97$ (diamonds), are shown in the core 
structure diagram of Fig.~\ref{cores_by_anchoring}. At high temperature the single core is stable for 
$w \lesssim 4.9$ while the split core is stable for $w \gtrsim 5.5$. The double core is stable for intermediate 
anchoring strengths $w$. The transition between these core structures occurs at lower values of $w$ and the 
surface core splitting decreases as $\tau$ decreases. When compared with the quadratic potential 
at the same temperature and colloidal radius, the transitions for the quartic potential occur at lower 
$w$.

In summary, at fixed colloidal size and temperature, a strong quartic potential stabilizes the 
split core structure. A strong quadratic potential is capable of stabilizing the split core boojums, only
for LCs with positive elastic anisotropy. For LCs with negative elastic anisotropy the split core boojums
do not appear. In this article we address only two cases $\eta = 2, \eta = -\frac{1}{2}$. Presumably, there exist a positive threshold value of $\eta$ above which quadratic surface potential favours the split core boojums. 
The double core structure is an intermediate structure that is stable over a limited range of anchoring strengths, 
for both types of surface potential.

\subsubsection{Temperature \\}
 
In the previous section  we have investigated the role of the anchoring strength on the stability
of the boojum core structures, and found evidence that the temperature plays 
a major role on the stability of the boojum cores. Here we consider the temperature as the 
control parameter of the structural boojum core transitions. 

 
We start by investigating a colloidal particle with a quadratic surface potential, $w_2=0$. 
We consider reduced temperatures in the range $-3.24\lesssim \tau \lesssim 1$
and set $R=0.1\mu m$, which is $\simeq 6.6 \xi $ in LCs with positive elastic 
anisotropy $\eta = 2$, and $\simeq 12.4 \xi$ in LCs with $\eta = -\frac{1}{2}$,
since the bulk correlation length $\xi$ depends on the value of $\eta$.  

\begin{figure}[t]
\centering
\includegraphics[width=0.45\textwidth]{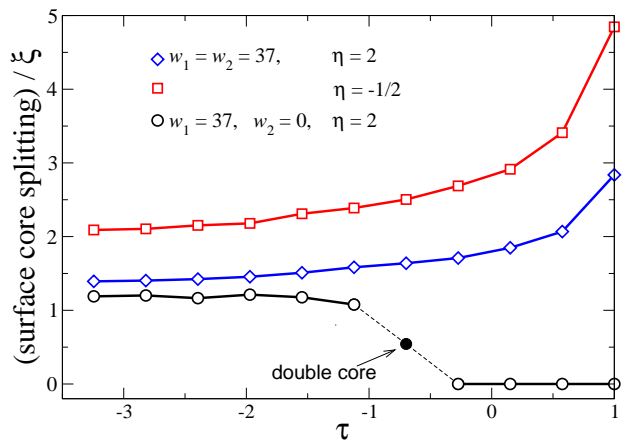}
\caption{Defect core transitions driven by the temperature. 
$R = 0.1 \mu m$ which is $\simeq 6.6 \xi $ for $\eta = 2$, and $\simeq 12.4 \xi$ for
$\eta = -\frac{1}{2}$. Split core boojums are stable for the quartic, $w_1$ and $w_2\neq 0$, potential
at all temperatures and for both $\eta=2$ (diamonds) and $\eta=-\frac{1}{2}$ (squares).
For the quadratic potential, $w_2=0$ and LCs with $\eta=2$ the sequence single-double-split cores is 
observed as the temperature decreases.} 
\label{cores_by_tau} 
\end{figure}

A quadratic potential ($w_2=0$) with $w_1\simeq 37$ stabilizes the split core structure 
at low temperatures only in LCs with positive elastic anisotropy, $\eta = 2$. The core 
structure diagram is depicted in Fig.~\ref{cores_by_tau} by circles. 
For $\tau \lesssim -1.12$ the split core structure is stable, while at high temperatures
$\tau \gtrsim -0.27$ the boojum has a single core structure. The transition between these
structures seems to occur continuously through the intermediate double core structure, which
 is observed in the range $-1.12\lesssim \tau\lesssim -0.27$. 
The surface splitting of the split core structure increases slowly as the temperature decreases and saturates
 at a value close to $\xi$.  

The split core structure was not observed in LCs with negative anisotropy, $\eta=-\frac{1}{2}$.
Recalling that $\eta>0$ corresponds to $K_1=K_3>K_2$, while $\eta<0$ to 
$K_1=K_3<K_2$, we note that unfavourable twist distortions are responsible
for the suppression of the split core structures in LCs with $\eta=-\frac{1}{2}$.
Twist deformations vanish when the azimuthal component of the director field, $n_\phi$,
vanishes \cite{stark1999}, which is the case for single core boojums. 
By contrast, for double and split core boojums $n_\phi$ acquires a non-zero value that 
increases from the double to the split core structures. 
Because the twist distortions are energetically unfavourable in LCs with $\eta<0$ 
the split core boojums are unstable with respect to the double core structure. 
This is in line with a twist transition that occurs at the core of a hyperbolic hedgehog defect 
(see Figs. 4, 5 in Ref.~\cite{stark1999}) on a colloidal particle with homeotropic anchoring, 
and in general, with the effect of elastic anisotropy on the structure of distortions 
around topological defects in $2D$ \cite{Silvestre.2009}. 


For a quartic potential with $w_1 = w_2 \simeq 37$ the split core is stable for both types of 
LC elastic anisotropy, at all temperatures. 
The core splitting increases with temperature up to $2.8 \xi$ for $\eta = 2$, and $4.9 \xi$ for $\eta = -\frac{1}{2}$, 
close to the NI coexistence temperature (see Fig.~\ref{cores_by_tau} diamonds and squares, respectively). The quartic 
term in the surface potential is quite effective in stabilizing the split core structure, even when twist distortions 
are unfarourable ($\eta = -\frac{1}{2}$). In fact, the split core is not only stabilized, but its splitting is even larger than that 
observed in systems with $\eta>0$. 
We note that the boojum size is proportional to the splitting parameter shown in Fig.~\ref{cores_by_tau} 
and it decreases as the temperature decreases.

In summary,  strong quadratic surface potentials and low temperatures 
favour split core boojums only for LCs with positive elastic anisotropy. 
When $\eta < 0$ the double core structure is stable. 
Strong quartic potentials favour split core boojums, which 
are the only stable structures, regardless of the LC anisotropy.

\subsubsection {Particle radius \\ }

Finally, we investigate the effect of the colloidal radius $R$ on 
the structural transitions of the boojum cores. For this purpose we consider colloids 
with sizes in the range $0.05 \mu m  \leqslant R \leqslant  0.5 \mu m$. 

\begin{figure}[t]
\centering
\includegraphics[width=0.45\textwidth]{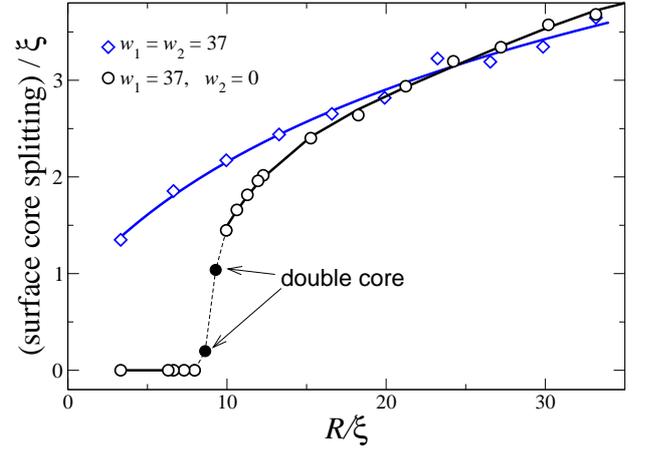}
\caption{Defect core transitions driven by the colloidal size, for quadratic (circles) 
and quartic (diamonds) potentials with $w\simeq37$, at $\tau \simeq 0.16$. 
For quadratic potentials, single core boojums are stable at small and spilt 
core boojums are stable at large colloidal sizes, $R$. The LC anisotropy $\eta = 2$. 
For quartic potentials, the split core structure is stable over the whole range of $R$.
The solid lines are guides to the eye.} 
\label{cores_by_R} 
\end{figure}


Boojum core transitions driven by the colloidal size, for quadratic (circles) and quartic 
(diamonds) potentials are illustrated in Fig.~\ref{cores_by_R}. 
For LCs with $\eta = 2$ the quadratic potential stabilizes the 
split core boojum on large colloids, $R/\xi \gtrsim 10$, the single 
core boojum on small colloids, $R/\xi \lesssim 8$, and the double 
core boojum at intermediate colloid sizes. By contrast, in LCs with 
$\eta=-\frac{1}{2}$, and the same quadratic potential, the split core is not 
observed and the single-double core transition occurs at $R/\xi \simeq 45$ 
(not shown). This behaviour is similar to the one described in previous sections, 
for quadratic potentials. When twist distortions are energetically unfavourable 
they are suppressed rendering the split core structure unstable.
For a quartic potential with $w_1 = w_2$ the split core is stable for both types 
of elastic anisotropy in the whole range of colloidal sizes. The size of the half-ring, 
given by the core splitting on the colloidal surface, increases with $R$ and saturates at $\simeq 6 \xi$ 
for $\eta = 2$, and at $\lesssim 10 \xi$ for $\eta = -\frac{1}{2}$. Again, the quartic surface 
potential Eq.~(\ref{Fournier}) stabilizes split core boojums in LCs with negative 
elastic anisotropy.

We end this section by noting that although, it may be difficult to observe the reported
boojum core structures using standard optical methods, their influence on the effective
interaction between colloidal particles at short distances is significant, as will be discussed in subsection \ref{3D_two_colloids}. The far-field nematic configuration, however,
is not affected by the mesoscopic structure of the boojum cores.

\subsection{Two spherical particles}
\label{3D_two_colloids}

\begin{figure}[t]
\centering
\includegraphics[width=0.25\textwidth]{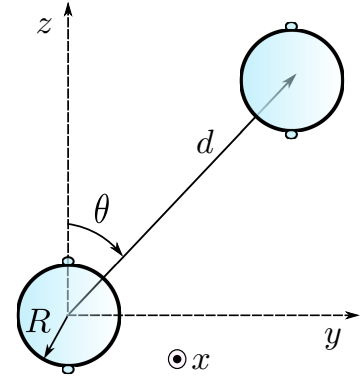}
\caption[2Spheres]{Schematic representation of two interacting boojum-colloids. 
The far-field director is parallel to the $z-$axis. For large $d$
pairs of boojums are aligned parallel to the far-field director.
\label{2_colloids} 
}
\end{figure}

In this section we shall calculate the two-body effective interaction potential $F^{(2)}(d,\theta)$
\cite{tasinkevych2006} for colloids with a quadratic surface potential, Eq. (\ref{Fournier}).
For the definition of $d$ and $\theta$, see Fig.~\ref{2_colloids}. $F^{(2)}$ is defined by decomposing the excess 
(over the free energy of the uniform nematic) free energy $F_2\equiv F_{\mathrm{LdG}} - f_b(Q_b)\Omega$ for
the system of two particles as follows
\begin{equation}
 F_2(d,\theta)= 2 F^{(1)} + F^{(2)}(d,\theta).
\label{pair_pot_def}
\end{equation}
Here $F^{(1)}$ is the excess free energy calculated independently for one isolated colloid.
By definition $F^{(2)}$ tends to zero when $d\rightarrow \infty$.

 In particular, we shall demonstrate how the non-linear effects become dominant, leading to a qualitative change of 
the effective interaction, e.g.,  the potential $F^{(2)}(d,\theta=0)$ changes from repulsive at large 
$d$ to strongly attractive at moderate to short distances. As we shall see this is driven by the 
complex behaviour of the interaction between the topological defects that results in a re-arrangement 
of their structure at short distances. Analogous changes in the effective interaction have been 
reported experimentally~\cite{Smalyukh.2005a} and confirmed within LdG theory~\cite{Mozaffari.2011} 
where it has been shown that the axial symmetry of the location of the boojum pairs is continuously broken, 
resulting in an effective colloidal attraction. 

Here we shall re-examine the effective colloidal interaction based on the results of the previous section 
where we have shown that the axial symmetry of the boojum cores may be broken 
at the level of a single colloidal particle. In what follows we use the notation set in 
Fig.~\ref{2_colloids}. The far-field director is aligned with the $z$-axis. Due to the 
spherical symmetry of the colloids,  $F^{(2)}$ depends on the colloidal separation $d$ and on the polar angle 
$\theta$ only. In this section we consider colloidal particles of equal size, $R=0.1\mu m$,
the anisotropy of the elastic constants $\eta=2$, the reduced temperature $\tau\simeq 0.16$,
and the other model parameters corresponding to the LdG parametrisation of 5CB,  described in 
subsection ~\ref{LdGmodel}. For this set of parameters, the boojum cores exhibit a single core 
configuration over the whole range of anchoring strength $w\equiv w_1$.
Ultimately, we aim at explaining the experimental results of Smalyukh et al. 
\cite{Smalyukh.2005a}, namely the force versus $\theta$ curves, which disagrees strongly with 
the asymptotic quadrupole-quadrupole result, at moderate and short colloid-colloid distances. 

\begin{figure}[t]
\centering
\includegraphics[width=0.45\textwidth]{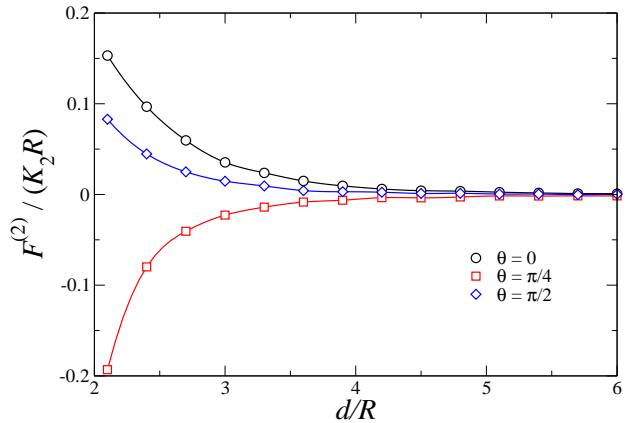}
\caption[Spheres]{Effective interaction potential $F^{(2)}(d)$ for several values of $\theta$. 
The surface potential is quadratic, Eq. (\ref{Fournier}), and the anchoring strength is 
$w\equiv w_1\simeq0.37$. Elastic anisotropy $\eta=2$, radius of the particles $R\simeq 6.6 \xi$, and
reduced temperature $\tau\simeq 0.16$. 
 \label{interact_W_0.0001} 
}
\end{figure}

In Fig.~\ref{interact_W_0.0001} we plot $F^{(2)}(d)$ in the weak anchoring 
regime. In this regime the interaction is repulsive when the inter-colloidal vector 
is either parallel or perpendicular to the far-field director, $\theta=0$ or $\pi/2$ respectively, 
and is attractive at intermediate colloidal orientations, as expected for quadrupole-quadrupole interactions. 

\begin{figure}[t]
\centering
\includegraphics[width=0.5\textwidth]{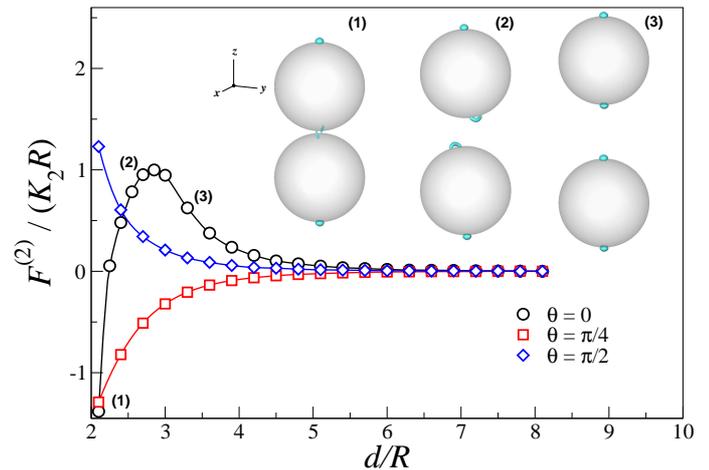}
\caption[Spheres]{Effective interaction potential $F^{(2)}(d)$  for several values of $\theta$. 
The anchoring strength is $w\simeq 37$. Elastic anisotropy $\eta=2$, radius of 
the particles $R\simeq 6.6 \xi$, and reduced temperature $\tau\simeq 0.16$.
The insets depict the boojum configurations for several distances $d$ and $\theta = 0$. In 
the attractive range, $2R\lesssim d\lesssim 3R$, the inner boojums are $\frac{1}{2}$ 
disclination rings represented by the iso-surface of the scalar order-parameter $Q = 0.3 Q_b$. 
 \label{interact_W_0.01} 
}
\end{figure}

The character of the effective interaction changes significantly in the strong anchoring regime, as shown 
in Fig.~\ref{interact_W_0.01}. For oblique, $\theta=\pi/4$, and perpendicular, $\theta=\pi/2$, orientations of the colloidal pair the interaction remains qualitatively the same: attractive when the colloids are at an oblique angle with the far-field director and purely repulsive when the colloids are perpendicular to it. However, at $\theta=0$ the effective interaction is no longer purely repulsive. At short distances, $d\lesssim3R$, the particles start to attract each other. This change results from a symmetry break of the boojum-pair configuration as shown in the insets of Fig.~\ref{interact_W_0.01}. At large distances the boojums are located at the poles of the particles, see  inset (3) in Fig.~\ref{interact_W_0.01}, and aligned parallel to the far-field director, as in the case of an isolated particle. As the separation between the colloids decreases, the repulsion of the inner defects increases. The change from repulsive to attractive colloidal interaction 
is driven by a change in the position of the inner defects, see inset (2) in Fig.~\ref{interact_W_0.01}. 
This mechanism for attractive colloidal interactions -- the re-arrangement of the inner defects 
with quadrupolar symmetry -- was reported almost 10 years ago for $2D$ colloids with homeotropic 
anchoring~\cite{Tasinkevych.2002}. We note that although in the single particle configuration the 
boojums are in the single core state, the inner boojums undergo a transition to the split core 
state as the distance between the colloidal particles decreases, see the inner boojums in 
Fig.~\ref{interact_W_0.01}(2). The symmetry break in the position of the inner boojums is accompanied 
by a symmetry break of the inner boojums core structure, which changes from the axially symmetric single 
core to the asymmetric split core configuration. The global minimum of $F^{(2)}(d,\theta)$
corresponds to close contact, $d = 2R$, at an oblique angle $\theta$ close to $\pi/4$. This is in 
line with the experimental results of \cite{Poulin.1998, Smalyukh.2005a} where the boojum-colloids 
were reported to coalesce. In Sec.~\ref{2D} we shall show that this is not the case for boojum-colloids in $2D$ systems.

\begin{figure}[t]
\centering
\includegraphics[width=0.5\textwidth]{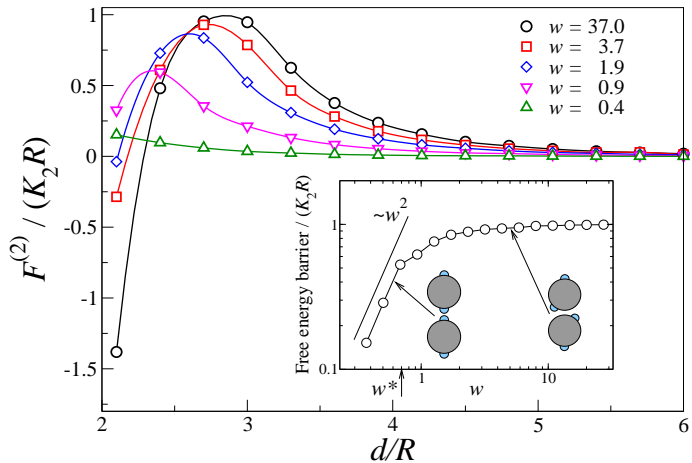}
\caption[Spheres]{Effective interaction potential as a function of the distance $d$ between two 
spherical colloidal particles, aligned with the far-field director, $\theta=0$. The surface anchoring
potential is quadratic, Eq. (\ref{Fournier}). Elastic anisotropy $\eta=2$, radius of the particles $R\simeq 6.6 \xi$, and
reduced temperature $\tau\simeq 0.16$. The anchoring strength $w$ varies in the interval $[0.4 , 37]$. 
The inset depicts the height of the free energy barrier as a function of $w$, 
exhibiting quadratic behaviour $~w^2$ in the weak anchoring regime.
 \label{pair_theta_0_barrier_vs_W} 
}
\end{figure}

The effect of the anchoring strength $w$ is trivial for colloidal orientations $\theta = \pi/4, \pi/2$, 
and thus we focus on  $\theta = 0$. In Fig.~\ref{pair_theta_0_barrier_vs_W} 
we plot $F^{(2)}(d,\theta=0)$ for several values of anchoring strengths $w$. The dependence of the free energy 
barrier on the anchoring strength is illustrated in the inset. The barrier height is weakly discontinuous at $w = w^*\simeq 0.7$ delimiting two types of behaviour. For strong anchoring, $w>w^*$, the free energy at distances close to the free energy 
maximum, $d^*$, is a smooth function of $d$. This implies that as the colloidal separation decreases the boojums 
rearrange smoothly, driving the change from repulsive to attractive interactions. For weak anchoring, 
$w<w^*$, the nematic order at the poles of the particles is suppressed, but there are no defects.
The system behaves almost linearly in this regime, i.e., ${\mathbf Q} \propto w$ and the resulting 
free-energy $\propto w^2$. The effective interaction is repulsive at all $d$, as expected for two 
quadrupoles oriented at $\theta = 0$.

We conclude that the interaction between colloids aligned with the far-field director changes from 
repulsive to attractive at short-distances as the result of the re-arrangement of 
the positions and of the  core configurations of the inner boojums. 

Finally, we shall give a detailed description of the angular dependence of the effective force acting on the colloids. 
The experimental measurement of this force has been reported in Ref.~\cite{Smalyukh.2005a}. According to this reference
brownian motion dominates the dynamics for $d>6R$, but at shorter distances ($d\lesssim4R$) the dynamics is governed 
by the nematic mediated force. In this regime the force of the order of the elastic constant $K$ has been found.
\cite{Smalyukh.2005a}. Since the measured elastic force was significant only for relatively small distances, 
strong deviations from the asymptotic quadrupolar interaction have been reported. For instance, the 
orientation $\theta$ corresponding to the strongest attraction has been found to depend on the distance between colloidal 
particles, while the interaction of quadrupolar type exhibits maximal attraction at $\theta \approx 49^{\circ}$, for any $d$. In addition, the attraction for  $0^o<\theta < 70^o$ and repulsion for  $75^o < \theta < 90^o$ has been found, 
while quadrupoles repel at $\theta = 0$. Finally, the measured radial dependence of the 
force (at fixed $\theta = 30^o$) deviated systematically from the quadrupolar power law $\propto d^{-6}$ 
for $d \lesssim 3R$. These findings suggest that non-linear effects dominate and that the 
superposition approximation does not hold in the range of distances that have been
investigated.
 
We have calculated the absolute value $\mathrm F$ of the elastic force,
${\mathbf  F} = {\mathrm F}_r {\mathbf e}_r + {\mathrm F}_{\theta} {\mathbf e}_{\theta}$, 
as well as its radial and polar components ${\mathrm F}_r = -\partial F_{\mathrm{LdG}} / \partial d$ 
and ${\mathrm F}_{\theta} = -(1/d) \partial F_{\mathrm{LdG}} / \partial \theta$
by numerical differentiation.
\begin{figure}[t]
\centering
\includegraphics[width=0.45\textwidth]{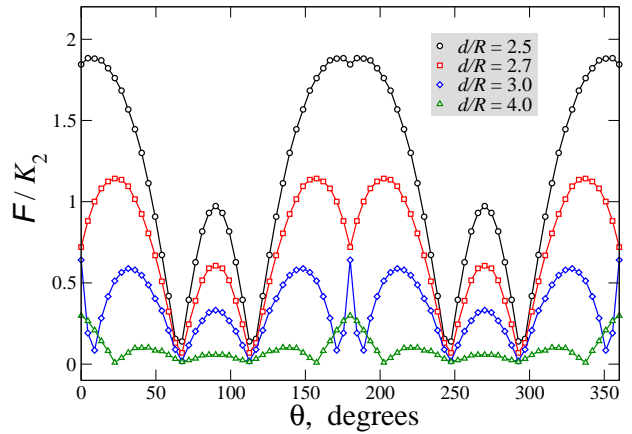}
\caption[Spheres]{Absolute value of the pairwise force as a function of  $\theta$,
for several values of $d$.  $\eta =2$, $R  \simeq 6.6\xi$, $\tau \simeq 0.16$ and 
anchoring strength $w \simeq 37$. 
 \label{force} 
}
\end{figure}
In Fig.~\ref{force} the resulting magnitude of the force is plotted as a function of the relative orientation, 
$\theta$, for several values of the  particles separation $d$. We note that as the separation between the particles increases the magnitude of the force decreases and its structure as a function of $\theta$ decreases. In Fig.~\ref{force} 
the angular dependence of the force, at separation $d=4R$, resembles that of the quadrupolar force 
\cite{Ruhwandl.1997} with principal maxima at $\theta=0$ and $\theta=\pi$. In addition, the force exhibits 
three secondary maxima at intermediate values of $\theta$ in line with the angular dependence of the quadrupole-quadrupole 
force \cite{Ruhwandl.1997,Smalyukh.2005a}. As the distance between the colloids decreases the angular dependence of the force 
changes quite rapidly and quite drastically. Just below $d=3R$ the maximum at $\theta=\pi$ becomes a minimum (the force weakens) 
and the difference between the secondary maxima at intermediate values of $\theta$ becomes more pronounced.
A similar angular dependence was observed in the force reported in Ref.~\cite{Smalyukh.2005a} at colloidal separations 
$3R \leqslant d \leqslant 4R$. However, our theoretical results appear to be more sensitive to the inter-colloidal 
separation and predict different angular behaviours at colloidal separations $d=3R$ and $d=4R$ at $\theta=0$. 
At this orientation the experiments report a local minimum for colloidal separations in the range $3R \leqslant d \leqslant 4R$ 
\cite{Smalyukh.2005a}.

\begin{figure}[t]
\centering
\includegraphics[width=0.45\textwidth]{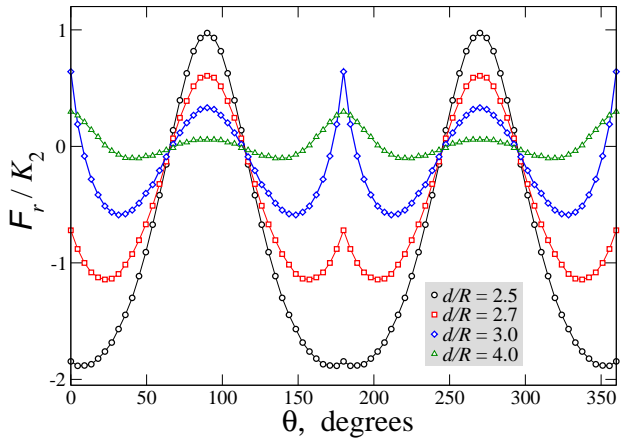}
\caption[Spheres]{The radial component of the force between two colloidal particles as a function of  $\theta$,
for several values of $d$. The values of the model parameters are the same as in Fig.~\ref{force}.
\label{force_r} 
}
\end{figure}

\begin{figure}[t]
\centering
\includegraphics[width=0.47\textwidth]{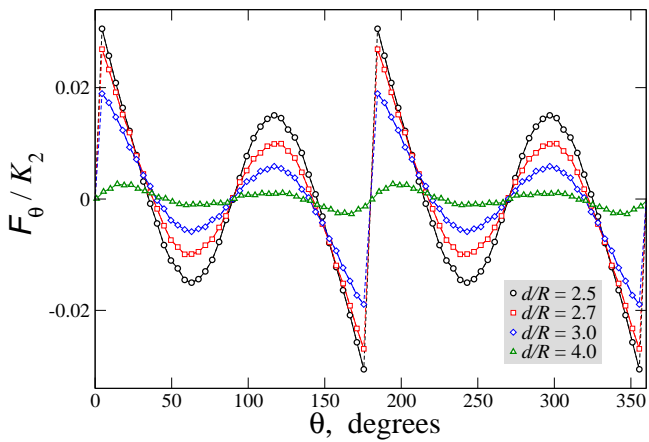}
\caption[Spheres]{The polar component the force between two colloidal particles as a 
function of the colloidal orientation, $\theta$, at several values of the particles separation 
$d$. The values of the model parameters are the same as in Fig.~\ref{force}.
 \label{force_theta} 
}
\end{figure}

In an attempt to rationalise the discrepancies between the theoretical and the experimental results we 
have plotted the radial and polar components of the force as a function of $\theta$ in Figs. \ref{force_r} and \ref{force_theta}.
The radial component of the force is found to be two orders of magnitude larger than the polar component, 
and thus it dominates the behaviour of the total force. This is in sharp contrast with the measured force, 
which is characterised by a radial component that differs significantly from the total force at short distances, 
around $\theta=0$ (see Fig. 4 of reference \cite{Smalyukh.2005a}, in particular the inset c) where the radial 
component of the force at $d=3R$ appears to vanish at $\theta=0$). The results for the polar component of the 
force plotted in Fig. \ref{force_theta} reveal that the angular dependence changes abruptly at $\theta=0$, by contrast to 
the behaviour of the force at $\theta=\pi/4$ and $\pi/2$. This is related to the head-on interaction of the intermediate 
boojums and the associated discontinuity of ${\mathrm F}_{\theta}$ at $\theta =0$. This, probably, has had an impact 
on the experimental results as the colloidal configuration is unstable at $\theta = 0$ rendering an accurate force measurement, at this
orientation, very difficult. Finally, we note that the colloidal particles in the experiment are of the order of 
$2\mu m$, while we considered colloids one order of magnitude smaller. This may also account for some discrepancies 
between the theoretical and experimental results.

\section{Two-dimensional systems}
\label{2D}

Dispersed colloidal particles in smectic $C$ (sm$C$) films are realised either as  inclusions of a lower order phase 
(isotropic, nematic, or cholesteric) or as smectic islands with a higher number of layers than the surroundings \cite{Dolganov.2006,Bohley.2008}. Planar boundary conditions at the colloidal particle may lead to the nucleation
of a pair of surface defects  \cite{Cluzeau.2002,Dolganov.2006}.
Planar anchoring may also lead to the nucleation of a satellite (bulk) defect \cite{Silvestre.2009}, which will not be considered here. At large distances the interaction between these particles is of quadrupolar type, and deviates from it
at short-distances. By contrast to the observations in 3D systems, where the spherical colloids with planar anchoring come into contact and coalesce \cite{Poulin.1998}, inclusions in sm$C$ films maintain a well-defined separation $d_{eq}\simeq 2.7R$, where $R$ is the inclusion radius. 

The  distortions in sm$C$ films due to an isolated inclusion with planar anchoring, as well as the asymptotic effective pair interaction, have been described by invoking the electrostatic analogy and the superposition approximation. In the simplest approach \cite{Cluzeau.2005} the distortions around an inclusion are modelled by the superposition of three solutions to the Laplace equation  corresponding to one topological defect with charge $+1$ in the center of the inclusion and two $-\frac{1}{2}$ surface defects. Then by using the superposition approximation the quadrupole-quadrupole effective interaction is obtained. However, this solution does not satisfy the boundary conditions at the inclusion boundary, and the superposition approximation fails to describe the repulsion observed experimentally at small distances \cite{Cluzeau.2005}. A related approach \cite{Silvestre.2008}, based on the exact solution for the ${\bf c}-$director (defined below) around an inclusion with arbitrary anchoring strength \cite{Burylov.1994} reveals that the repulsion appears simultaneously with the nucleation of the defects. Nevertheless, neither approach describes the non-linear effects responsible for the re-orientation of the defects at short distances. 

In what follows, we shall use the Landau description of sm$C$ films in order to calculate the interaction between two inclusions with planar anchoring. We aim at a detailed description of the defect structure, and we will show that the short distance colloidal interactions result from the re-arrangement of the defect positions, as in other $2D$ and $3D$ systems with quadrupolar symmetry \cite{Tasinkevych.2002,Mozaffari.2011}.

\subsection{Free energy of smectic $C$ films}

\begin{figure}[t]
\centering
 \includegraphics[width=0.3\textwidth]{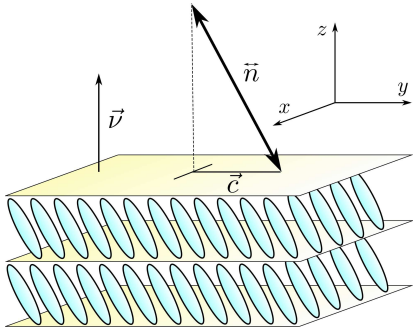}
 \caption{Schematic representation of a smectic $C$. $\mathbf{\nu}$ is the normal to the layers, $\nvec$ is the average local molecular orientation, and $\cvec$ is the in-plane projection of $\nvec$.} 
\end{figure}

In sm$C$ phases the LC molecules are organized into $2D$ layers with the local average molecular orientation $\nvec$ at a characteristic angle $\phi$ with respect to the layer normal $\mathbf{\nu}$. When layer deformations are negligible there are no variations of $\phi$ in the direction of $\mathbf{\nu}$, $\nabla_\perp\phi=0$. In this case, the relevant order-parameter is the in-plane projection of the director $\nvec$. This is an ordinary (variable length) $2D$ vector field $\cvec$. 

The simplest mesoscopic model for sm$C$ films is obtained by expanding the free energy in terms of the invariants of the order-parameter $\cvec$ and its derivatives $\partial_i c_j$ \cite{Silvestre.2006}. The free energy is then written as $F=l \int_{\bar{\Omega}} \left(f_{b}+f_{e}\right)\,\dd^2x$, where $l$ is the thickness of the film, and the bulk and elastic free energy densities are, at lowest order,
\begin{eqnarray}
f_b &=& -\frac{{\bar a}(T)}{2}|\cvec|^2 + \frac{{\bar b}}{4}|\cvec|^4, \label{smC_bulk1} \\
f_e &=& \frac{\bar {K}}{2}\left(\left(\nabla\cdot\cvec\right)^2 + \left(\nabla\times\cvec\right)^2\right),
\end{eqnarray} 
where, for simplicity, we assume the one-elastic constant approximation ($\bar {K}_1=\bar {K}_3=\bar {K}$). 
$\bar{a}(T)$ is assumed to be a linear function of temperature $T$ and $\bar{b}$ is a constant. 
Dimensionless free energy densities (the free energy $F$ has units of volume) are defined through the rescaling $\tilde{\cvec} = \sqrt{\bar{b}/(2\bar{a})}\cvec$ and $\tilde{F}=(\bar{b}/\bar{a}^2)F$, 
\begin{eqnarray}
 \tilde{f}_{b} &=& |\tilde{\cvec}|^2 \left(|\tilde{\cvec}|^2-1\right) \label{SmCbulk}\\
 \tilde{f}_{e} &=& \xi^2\left(\left(\nabla\cdot\tilde{\cvec}\right)^2 + \left(\nabla\times\tilde{\cvec}\right)^2\right),
\label{SmCelastic}
\end{eqnarray} 
where $\bar{\xi}=\sqrt{\bar{b} \bar{K}/\bar{a}(T)^2}$ is the correlation length. Eq.~(\ref{SmCbulk}) predicts non-zero values for the bulk scalar order-parameter $|\tilde{\cvec}_b|=1/\sqrt{2}$ at all temperatures. 
Topological defects correspond to regions of reduced orientational order $|\tilde{\cvec}|\to 0$, 
and their topological charge is defined by the winding-number of the vector field $\tilde{\cvec}$ \cite{Gennes.1993}.

Due to the vector nature of the $\cvec$ director, it is possible to distinguish clockwise, anticlockwise or mixed planar anchoring on circular inclusions in sm$C$ films \cite{Silvestre.2009}. The orientation that is realised results from the interplay of the anchoring potential, the $\cvec$ director field in the vicinity of the inclusion, and the LC properties. We consider the case of mixed planar anchoring which favours the nucleation of boojum pairs \cite{Silvestre.2008}, the $2D$ counterpart of the $3D$ boojums investigated in subsection~\ref{single}. We use a surface potential of the Rapini-Papoular form \cite{PAPOULAR.1969} which is the $2D$ version of the quadratic surface potential, 
Eq.~(\ref{Fournier}),
\begin{equation}
F_s=l \int_{\partial\bar{\Omega}}{ \frac{W}{2}\left(\frac{\cvec\cdot\nu}{|\cvec_b|}\right)^2\dd l}.
\end{equation}
where $W$ is the anchoring strength, and $\mathbf{\nu}$ is the normal to the surface $\partial\bar{\Omega}$ of the inclusion. We note that in $2D$ quartic terms do not change the surface potential qualitatively and will not be considered. Rescaling the variables we obtain
\begin{equation}
\tilde{F}_s=l \int_{\partial\bar{\Omega}}{ \frac{w\xi^2}{2R}\left(\frac{\tilde{\cvec}\cdot\nu}{|\tilde{\cvec}_b|}\right)^2\dd l},
\label{Rapini-Papoular}
\end{equation}
 where we introduced the dimensionless anchoring strength $w=WR/\bar{K}$. 

\subsection{Single circular particle}
\label{single_incl2D}

In this section we consider a single circular inclusion in a sm$C$ film. The mixed planar anchoring is enforced by fixing the far-field $\cvec-$director parallel to the $y-$axis. In this case a pair of boojums may nucleate at the particle surface. Within the FO elastic theory the $\cvec-$director is a unit vector $\cvec=(\cos{\Phi},\sin{\Phi})$. The analytic solution for the orientational field, $\Phi$, at arbitrary anchoring  strength $w$, can be written as \cite{Burylov.1990,Burylov.1994}
\begin{equation}
\Phi(r,\varphi)=-\arctan\left[\frac{\left(R/r\right)^2 p(w)\sin 2\varphi}{1-\left(R/r\right)^2 p(w)\cos 2\varphi}\right],
\label{Phi}
\end{equation}
where $r,\varphi$ are polar coordinates, and $p(w)=(2/w)\left(\sqrt{1+(w/2)^2}-1\right)$. 
The corresponding free energy is $F_{FO}=\pi l \bar{ K} \left(-\log\left(1-p^2\right)+\left(w/2\right)\left(1-p\right)\right)$. 
The solution, Eq.~(\ref{Phi}), describes non-singular surface disclinations with a core size $r_c = (8 R/w^2) \Bigr ( \sqrt{1 + w^2/4}  -1\Bigl )$ \cite{Burylov.1994}, which 
for strong anchoring behaves as $r_c \simeq 4R / w$. For any finite $w$ 
the elastic free-energy is also regular, with the leading large $w$ behaviour $F(w \gg 1)\simeq \pi \bar{K} l \ln (w)$. Singular topological defects appear only in the limit $w\rightarrow\infty$.

In the strong anchoring regime $w\rightarrow \infty$ the exact solution, Eq.~(\ref{Phi}), may be approximated by the sum of a finite number of solutions $\Phi_i$ to the Laplace equation, where $\Phi_i$ represents a point singularity (or topological defect) of some winding number $q_i$ \cite{Stark.2001}. The number of these topological defects, their locations, and winding numbers, are chosen to comply with the boundary conditions. The behaviour of the $\cvec-$director near the boojums suggests that these are  $-\frac{1}{2}$ topological defects. Since the colloidal particle with rigid planar anchoring accounts for a $+1$ virtual defect in the center, a naive guess would be to represent $\Phi$ as the sum of three point singularities, which we write as $(-\frac{1}{2}) + (+1) + (-\frac{1}{2})$ \cite{Cluzeau.2005}. Two $(-\frac{1}{2})$ singularities are located on opposite poles of the inclusion and are oriented along the far-field $\cvec-$director. However, a more careful analysis reveals that this ansatz violates the boundary condition at the surface of the inclusion, and additional virtual defects must be taken into account. The asymptotic $w\rightarrow\infty$ behaviour of the solution, Eq.~(\ref{Phi}), indicates that the correct ansatz corresponds to the triplet $(-1) + (+2) + (-1)$, where a $(+2)$ virtual defect is placed in the center of the inclusion and two $(-1)$ surface defects may be thought of as the sum of two $(-\frac{1}{2})$ defects, 
one of which is virtual and the other real \cite{Fukuda2007}.

\begin{figure}[t]
\centering
 \includegraphics[width=0.5\textwidth]{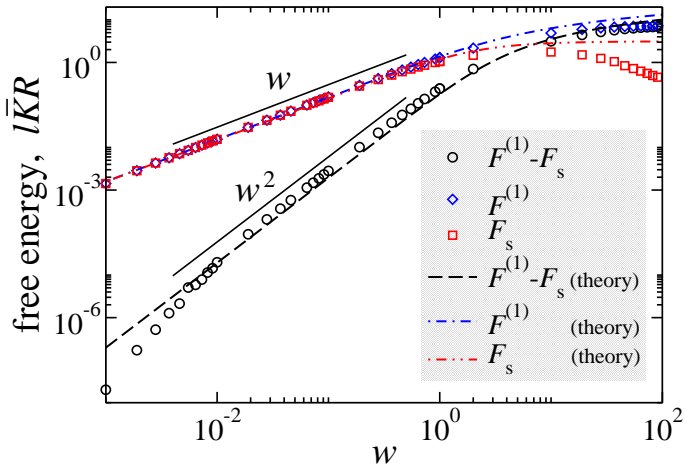}
 \caption{Elastic free energy $F^{(1)}-F_s$, excess (over the free energy of the uniform system) free energy $F^{(1)}$, and surface free energy $F_s$ as a function of the planar anchoring strength $w=WR/\bar{K}$. Points  correspond to the numerical minimization of the Landau free-energy functional (Eqs.~(\ref{SmCbulk}), (\ref{SmCelastic}), (\ref{Rapini-Papoular})). Dashed lines are obtained using the exact solution, Eq.~(\ref{Phi}). The radius of the particle  $R=100\bar{\xi}$.}
 \label{energy2Dsingle}
\end{figure}

In Fig. \ref{energy2Dsingle} we compare the excess free energy $F^{(1)}$, the surface free energy $F_s$, and the elastic free energy $F^{(1)}-F_s$, obtained by numerical minimization of the Landau free energy functional (Eqs.~(\ref{SmCbulk}), (\ref{SmCelastic}), and (\ref{Rapini-Papoular})) with the analytic results obtained by using the exact solution, Eq.~(\ref{Phi}), to the Frank-Oseen theory. Details of the numerical method may be found in the Appendix. In the weak anchoring regime the agreement is excellent. In the strong anchoring regime, however, the solutions start to deviate. Inspection of Fig.~\ref{energy2Dsingle} reveals that the discrepancies may be attributed to the differences in the surface free energies at large $w$. Indeed, the FO free energy is inaccurate in the presence of defects and ultimately diverges as $\sim\log(w)$. In the weak anchoring regime, $w\lesssim 1$, the excess free energy is dominated by the surface term which is linear in $w$, while the elastic free energy is quadratic, $w^2$, as in $3D$ systems, see Fig.~\ref{effect_of_W}.

\begin{figure}[t]
\centering
 \includegraphics[width=0.45\textwidth]{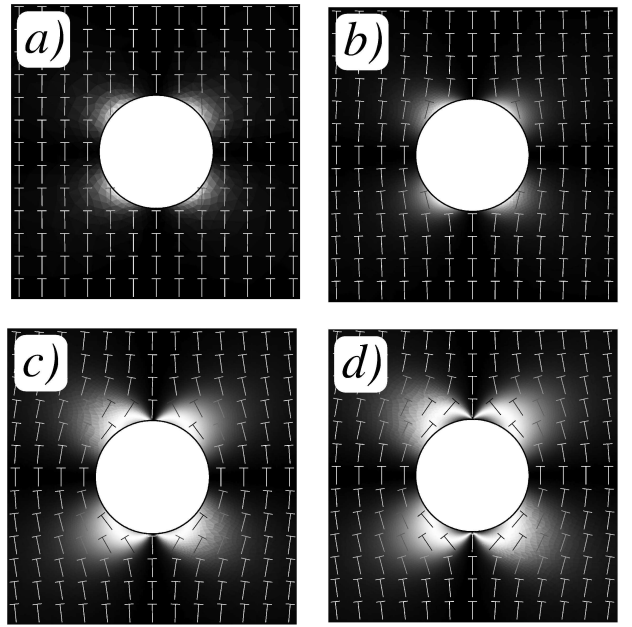}
 \caption{Equilibrium $\cvec-$director configurations for anchoring strengths, $a)$ $w=0.1$, $b)$ $w=1$, $c)$ $w=10$, and $d) w=100$. The grey-scale map corresponds to the light intensity $\propto \sin^2{2\Phi}$ as seen in experiments under cross-polarisers. The radius of the particle is $R = 100 \bar{\xi}$.}
 \label{single2D} 
\end{figure}

Figure \ref{single2D} illustrates the equilibrium $\cvec$-director configurations for increasing values of the anchoring strength $w$. The grey-scale map corresponds to the optical transmittance, $I/I_o\propto \sin^2(2\Phi)$, as seen in experiments under cross-polarisers. As the anchoring strength increases the inclusion-induced distortions, seen as bright fringes, grow. For $w>10$ a pair of topological defects nucleates at opposite poles of the particle (see Fig.\ref{single2D}$c)$ and $d)$). 
\begin{figure}[t]
\centering
 \includegraphics[width=0.45\textwidth]{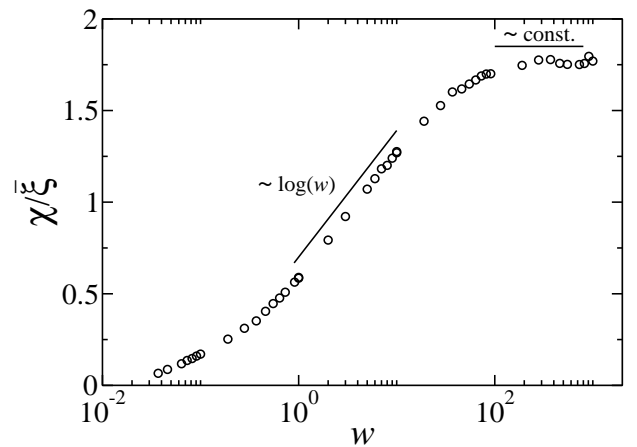}
 \caption{Effective radius $\chi$, Eq.~(\ref{chi_schopol}), of the 2D boojum core as a function of the anchoring strength $w=WR/\bar{K}$.}
 \label{schopohl} 
\end{figure}
In order to quantify the spatial extension of the boojums, we follow Ref.~\cite{Schopohl.1988} and introduce an effective core radius as follows
\begin{equation}
\chi=\left [\frac{1}{\pi}\int_\Omega{d^2x\left(1-\frac{|\cvec|}{|\cvec_b|}\right)}\right]^{\frac{1}{2}},
\label{chi_schopol}
\end{equation}
where the integral is over the region with one boojum. $\chi$  as a function of the anchoring  strength $w$ is plotted in Fig.~\ref{schopohl}. We distinguish three regimes. As the anchoring increases the LC molecules align along the preferred surface orientation creating small regions of low orientational order. At higher anchoring strength, $1\lesssim w \lesssim 10$, 
$\chi\propto\log(w)$, and the regions of low orientational order exhibit pre-nucleation of defects. Finally, in the nucleation regime, $w\gtrsim 10$, the effective radius increases with $w$ slowly, and eventually saturates at $\simeq 7/4 \xi$.

By contrast to the $3D$ case, the $2D$ boojums are always point-like and never split. This is a direct consequence of the topological constraint which forbids half-integer defects in sm$C$ films. In $3D$ nematics, however, half-integer ring disclinations are not only topologically allowed, but are even energetically favourable \cite{MORI1988,Penzenstadler.1989,sonnet.a:1995.a,Rosso1996,Mkaddem.2000}. This is one of the reasons why single core boojums in $3D$ may split into half-ring disclinations.

\subsection{Two circular particles}
\label{two2D}

In this subsection we shall analyse the pairwise effective interaction between inclusions with planar anchoring in sm$C$ films. At $d$ the inclusions ``decorated'' by boojums interact as $2D$ quadrupoles \cite{Tasinkevych.2002}
\begin{equation}
F_{quad}(d,\theta) \approx \frac{1-2\sin^2 (2\theta)}{d^4}.
\end{equation} 
This interaction is repulsive for $-\pi/8<\theta-m\pi/2<\pi/8$ and attractive for $\pi/8<\theta-m\pi/2<3\pi/8$, with $m=0,\,1,\,2,...$. 
At large distances the single-particle distribution of the $\cvec$-director is only  slightly perturbed by the presence of the other particle. As the separation decreases the perturbation increases and non-linear effects take over, changing the character of the interaction. The threshold distance $d^*$ at which the non-linear regime sets in depends strongly on the anchoring strength.

\begin{figure}[t]
\centering
\includegraphics[width=0.5\textwidth]{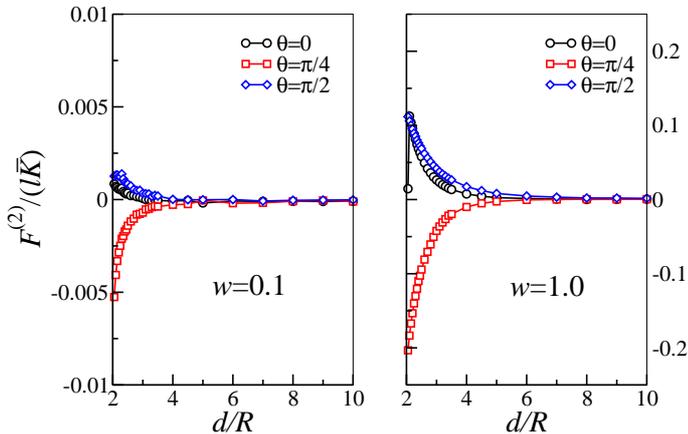}
\caption[Disks]{Interaction potential $F^{(2)}(d)$ for several values of $\theta$, anchoring strengths $w=0.1,\,1.0$ , and  radius $R = 100\bar{\xi}$.
 \label{2d_inclusions_1} 
}
\end{figure}

\begin{figure}[t]
\centering
\includegraphics[width=0.5\textwidth]{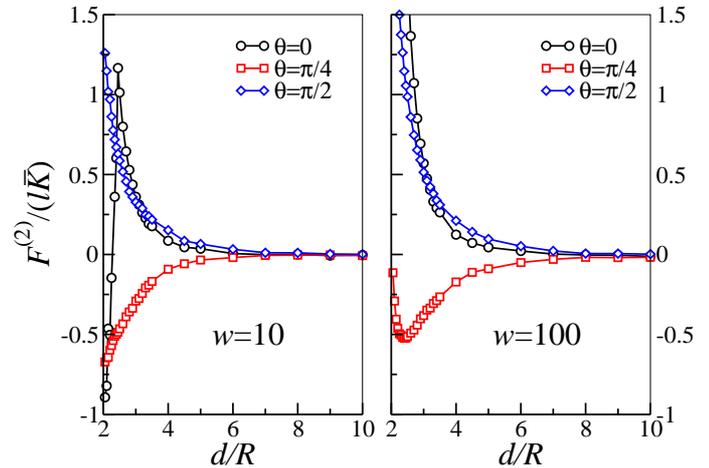}
\caption[Disks]{The same as Fig.~\ref{2d_inclusions_1}, but for anchoring strengths
$w=10,\, 100$. 
\label{2d_inclusions_2}
}
\end{figure}

\begin{figure}[t]
\centering
\includegraphics[width=0.45\textwidth]{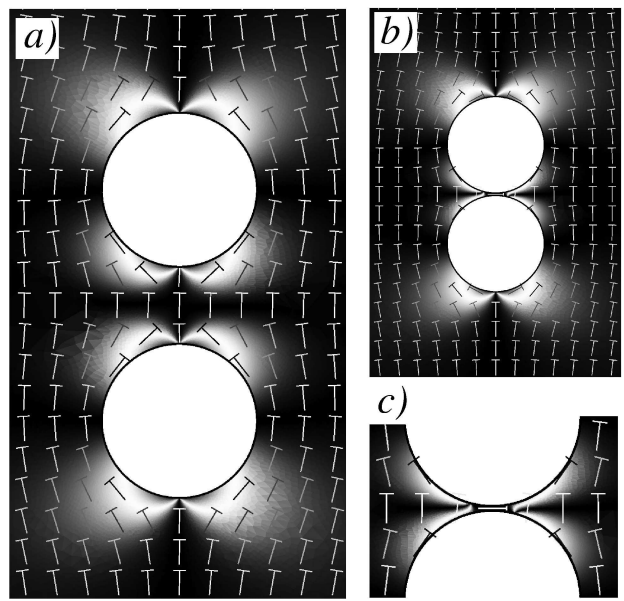}
\caption[Disks]{Equilibrium $\cvec-$director configurations for interacting colloidal particles with the anchoring strength $w=100$ at relative orientation $\theta=0$ and separations $a$) $d=3.0R$ and $b$) $d=2.05R$, $R = 100 \bar{\xi}$. $c$) close view of the configuration between the colloidal particles, for parameters as in $b$). The  greyscale map corresponds to the light intensity $\propto\sin^2 2\Phi$ as seen in experiments under cross-polarisers.
\label{2d_inclusions_3} 
}
\end{figure}

We minimize the Landau free energy functional, Eqs.~(\ref{SmCbulk}), (\ref{SmCelastic}), (\ref{Rapini-Papoular}) numerically. Details may be found in the Appendix. In Figs.~\ref{2d_inclusions_1} and \ref{2d_inclusions_2} we plot the effective interaction free-energy $F^{(2)}= F +F_s - F_0 -2F_{single}$ as a function of the particle separation, $d$, for anchoring strengths $w=0.1,\,1.0,\,10,$ and $100$. $F +F_s- F_0$ contains the contributions to the free-energy from the distortions of the $\cvec-$director,  $F_o\equiv F(\cvec_b)$ is the free energy of the uniform bulk system, and $F_{single}$ is the free energy of an isolated inclusion. At weak anchoring, $w=0.1$, $F^{(2)}$ is always repulsive for perpendicular $\theta = \pi/2$ or parallel $\theta = 0$ orientations, and attractive for oblique orientations, $\theta = \pi/4$.

When the anchoring strength increases regions of reduced order develop around the poles of the particles leading to an increase of the elastic free energy, see Fig.~\ref{energy2Dsingle}. This also results in an increase of the interaction strength at short distances, as can be seen in Fig.~\ref{2d_inclusions_1} for $w=1.0$ and Fig. \ref{2d_inclusions_2} for $w=10$. After the nucleation of topological defects the effective interaction changes drastically as shown in Fig.~\ref{2d_inclusions_2}, $w=100$, at separations $d \lesssim 4R$. 
In this regime the repulsion becomes even stronger for orientations $\theta=0,\,\pi/2$, but at $\theta = \pi/4$ the free-energy develops a well defined local minimum at $d\simeq 2.4 R$, which is followed by a repulsion for $d<2.4 R$. The latter prevents the inclusions from coalescing in agreement with the experimental observations \cite{Cluzeau.2002,Dolganov.2006}. We recall that this is in sharp contrast with the behaviour of $3D$ boojum-colloids, where no such repulsion is observed.

For strong anchoring and $\theta=0$, $F^{(2)}(d)$ changes from repulsive to attractive, with a discontinuous slope, at $d\simeq 2.45 R$, see Fig.~\ref{2d_inclusions_2}, $w=10$. Similar behaviour is observed for $w=100$ at $d\simeq 2.10 R$, and is not shown in Fig.~\ref{2d_inclusions_2} because the energy scale was chosen in order to emphasize the short range repulsion and the local minimum for $\theta=\pi/4$. Figure \ref{2d_inclusions_3} illustrates the equilibrium $\cvec$-director configurations for inclusions oriented along the far-filed $\cvec-$director, $\theta=0$, and for the separations $a$) $d=3.0R$, $b$) $d=2.05R$; the anchoring strength is $w=100$. For $d\gtrsim 2.10 R$ ($d\gtrsim 2.45 R$ for $w=10$) the inner boojums are aligned along the far-field $\cvec-$director, giving rise to an overall repulsion between the inclusions. As the separation decreases,
the inner boojums slide away from the $y-$axis in opposite directions, as shown in Fig.~\ref{2d_inclusions_3}$c$). This leads to a strong attraction between the particles. We emphasize that in $2D$ the re-arrangement of the inner boojums at $\theta = 0$ proceeds discontinuously, while in $3D$ -- continuously, see Fig.~\ref{interact_W_0.01} black circles. The contact value $F^{(2)}(d=2R,\theta=0)$ is, however, larger than $F^{(2)}(d\simeq2.4 R,\theta=\pi/4)$ which corresponds to the global minimum of $F^{(2)}(d,\theta)$ for the case of strong anchoring. 

\section{Conclusions}
\label{conclusions}

We have carried out a detailed study of the effective colloidal interactions for particles dispersed in a nematic host ($3D$) or in a smectic $C$ film ($2D$). In both cases planar (degenerate in $3D$) anchoring is imposed on the surfaces of the colloidal particles. Within the Frank-Oseen formalism in the strong anchoring regime, the boundary conditions are met by the creation of a pair of antipodal surface defects, boojums \cite{Mermin.1990}. For an isolated particle the vector connecting two boojums aligns itself with the far-field director, as required by the global boundary conditions \cite{Poulin.1998}. The resulting director field has quadrupolar symmetry, and the ensuing effective colloidal interactions exhibit quadrupole-quadrupole asymptotic behaviour \cite{Smalyukh.2005,Smalyukh.2005a,Cluzeau.2005,Dolganov.2006}, both in $2D$ and in $3D$. However, at short distances, where the superposition approximation fails and the non-linear effects dominate, the experimental results report significant deviations from this quadrupolar behaviour. The most important of these is the crossover from a repulsive to an attractive interaction at some threshold distance for particles aligned along the far-field director, $\theta =0$. The threshold distance is $\simeq3 R$ in $3D$, see Fig.~\ref{interact_W_0.01}, and $\simeq 2.5 R$ in $2D$, see Fig.~\ref{2d_inclusions_2}. This crossover is driven by the configurational reorientation of the inner boojums as the distance between the particles decreases. 

We have shown that the short-range colloidal interactions in $3D$ result not only from the re-arrangement of the defect positions, as predicted in $2D$ a decade ago \cite{Tasinkevych.2002} and described recently in $3D$ \cite{Mozaffari.2011}, but also from changes in the structure of the boojum cores. 
The description of the structural transitions between boojum cores is a challenging theoretical problem as distinct structures, characterized by tensor order-parameters that vary in regions of the order of the bulk correlation length, have similar free energies. Their stability results from a delicate balance of various contributions and structural transitions may be driven by the anchoring strength, the temperature or the colloid radius. We have concentrated on the mesoscale and addressed the structure and dynamics of the boojums and the resulting colloidal interactions based on the LdG free energy in $3D$ and $2D$, for spherically symmetric colloids. We have used finite elements methods with adaptive meshes in order to resolve the structure of the defect cores and establish the nature of the short-distance effective colloidal interactions. We have found that the defects are point like in $2D$, but acquire a rather complex structure in $3D$, which depends on the combination of the anchoring potential, the colloid radius, the temperature and the LC elastic anisotropy. In addition, we have found defect-core transitions between $i)$ single core, $ii$) double core, and $iii$) split core structures of the 
boojum pairs. 

Single core boojums are uniaxial along the $z-$axis (see Figs.~\ref{core_types}$a$, \ref{core_types_eigenvalues_along_z}$a$). The scalar order-parameter vanishes at one point above the colloidal surface, where the system is isotropic. Below this point the scalar order-parameter is negative and the nematic order is oblate. Above that point the scalar order-parameter is positive indicating that the nematic order is prolate. By contrast, double core boojums are biaxial along the $z-$axis (see Figs.~\ref{core_types}$b$, \ref{core_types_eigenvalues_along_z}$b$). The surface of maximal biaxiality intersects the $z-$axis at three distinct points separating regions of low nematic oblate and prolate order. Split core boojums are also biaxial along the $z-$axis. The surface of maximal biaxiality intersects the $z-$axis at two distinct points that delimit the region of oblate nematic order, Figs.~\ref{core_types}$c$, \ref{core_types_eigenvalues_along_z}$c$. As a general conclusion we find that on small colloids boojum cores are axially symmetric (see Fig.~\ref{cores_by_R}), point-like with index $+1$ -- the index referring to the charge of the projected director field on the surface --, which are stable at high temperatures (see Fig.~\ref{cores_by_tau}) and relatively weak anchoring (see Fig.~\ref{cores_by_anchoring}). On large colloids at low temperatures and strong anchoring, the axial symmetry is broken and the boojum $+1$ point-like cores split into a pair of $+\frac{1}{2}$ point-like surface defects, connected by a disinclination line (see Fig.~\ref{core_types}$c$). A structure without a fully developed disinclination line, the double core boojum, was also found (see Figs.~\ref{core_types}$b$, \ref{handled_hemisphere}). The detailed structure of the cores as well as the transitions between the different configurations depend in detail on the colloid and LC parameters.
We stress that the details of the core structure do not affect the far field configuration, i.e., on a spherical surface surrounding the boojum cores the order-parameter distribution resembles that of a point boojum. Their effect, however, on the short-distance interaction may be significant. Indeed, we have shown that the short-distance attraction at $\theta=0$, which results from the reorientation of the inner boojums, as the distance between the colloids decreases, corresponds to a structural change from single core to split core for the inner boojums pair.

We have shown how the non-linear effects become dominant, both in $2D$ and $3D$, leading to a qualitative change of the effective colloidal interaction, namely, the crossover from the large-distance quadrupole-quadrupole repulsion for a range of orientations around $\theta=0$, to an attraction at moderate to small distances, see Figs.~\ref{interact_W_0.01}, \ref{2d_inclusions_2}. This results from the complex dynamics of the topological defects that rearrange their position at short distances both in $2D$ and $3D$. This attraction, along the far-field direction, agrees with experimental observations \cite{Smalyukh.2005,Smalyukh.2005a,Cluzeau.2005,Dolganov.2006}, and with earlier calculations \cite{Mozaffari.2011} where it is shown that the axial symmetry of the boojum-pairs is broken when the repulsion changes into the attraction. The mechanism for attractive colloidal interactions -- driven by the re-arrangement of defects with quadrupolar symmetry -- was reported in $2D$ for homeotropic anchoring nearly a decade ago \cite{Tasinkevych.2002}. 

We emphasize an important difference between the $2D$ and $3D$ effective pair interaction potentials. In $3D$ the global minimum occurs at contact $d = 2 R$ for orientations at oblique angle with the far-field director. In $2D$ the optimal orientation of two inclusions is also oblique, but there is a free energy barrier $\simeq0.5 l \bar{K}$ keeping the particles apart. This distinction is in line with experimental observations in $2D$ and $3D$ \cite{Smalyukh.2005,Smalyukh.2005a,Cluzeau.2005,Dolganov.2006}.

Finally, we have calculated the force between boojum colloids in $3D$, see Fig.~\ref{force}. We have found, that as the distance decreases the angular dependence of the force changes rapidly and quite drastically. Just below $d=3R$ the maximum at $\theta=\pi$ becomes a minimum and the difference between the secondary maxima at intermediate values of $\theta$ becomes more pronounced. This angular dependence was actually observed by \cite{Smalyukh.2005a} at larger colloidal separations, $3R \le d \le 4R$. The theoretical results appear to be too sensitive to the inter-colloidal distance, when compared to the experimentally measured forces. 

\acknowledgments

We acknowledge partial financial support by  FCT-DAAD Transnational Cooperation Scheme
under the grant  N: 50108964,    
   FCT grants PEstOE/FIS/UI0618/2011, PTDC/FIS/098254/2008 and PTDC/BPD/50327/2007 (NMS),   
  FP7 IRSES Marie-Curie grant PIRSES-GA-2010-269181.

\appendix
\section{Numerics}
\label{appendix}

In this section we discuss briefly how  the scheme for numerical  minimization of the free energy functionals is implemented. In order to discretize the continuum models we resort to a finite element method \cite{wait_book} with adaptive meshes. For $3D$ systems the nematic and the colloidal particles of radius $R$ are confined in a cubic  box of linear size $30\times R$. For smectic $C$ films the system is confined in a $10 R \times 10 R$ square box.  

\subsection{Meshing}

The surfaces of the spherical colloidal particles are discretized using the open source \textit{GNU Triangulated Surface Library} \cite{GTS}. The library uses a recursive subdivision algorithm in order to triangulate the surface of a unit sphere. Starting from the icosahedron  as the first approximation the next levels of refinement are constructed by subdividing each triangle into four smaller triangles.
We have used the sixth refinement level corresponding to a surface mesh with $20480$ triangles. 

In the next step the domain $\Omega$ accessible to the nematic is triangulated. For $3D$ systems the triangulation is carried out using the \textit{Quality Tetrahedral Mesh Generator}, TetGen, \cite{tetgen} library. TetGen generates boundary conforming Delaunay \cite{delanay:1934} meshes of $3D$ domains with a piecewise linear boundary. The boundary can also be specified as a surface mesh, which is the case for the triangulated surfaces of the spherical particles. TetGen supports an isotropic size conforming triangulation, i.e., the resulting elements conform to a given local size map. For $2D$ systems the corresponding nematic domains are triangulated using the \textit{INRIA}'s $BL2D$ software \cite{BL2D}. The $BL2D$ package supports the creation of anisotropic meshes, where the triangulation process is governed by specifying a metric map.
 
\subsection{Minimization}

We use linear elements both in $2D$ and $3D$, i.e. the values of the nematic field ${\bf Q}_i$ 
are specified only at the vertices ${\bf x}_i$ of the mesh and a linear interpolation is used in order
to determine the values of ${\bf Q}(\bf x)$ at some other point ${\bf x}$ of $\Omega$.  
Due to the discretization
the volume and surface integrals in $3D$ are replaced  by sums of integrals 
over tetrahedral and triangular elements, respectively, and in $2D$  
by sums of integrals over triangular and line elements (the discretized perimeters
of the colloids). Each integral is evaluated numerically using generalized
Gaussian  quadrature rules for multiple integrals \cite{cub_encycl}.
For integration over tetrahedra a fully symmetric cubature rule with 11 points,
 which is exact for polynomials of degree 4,
is used \cite{cub_encycl,keast1986}.  
For integration over triangles a fully symmetric quadrature rule with 7 points,
which is exact for polynomials of degree 5 \cite{stroud1971}, is used. Finally, 
one-dimensional integrals are evaluated using the 5 points Gauss-Legendre quadrature, which is
exact for polynomials of degree 9. 

During the minimization the values of the tensor order-parameter ${\bf Q}$ at the vertices of 
the bounding box are kept fixed, equal to the values of the uniaxial bulk nematic. 
The values of ${\bf Q}_i$ at all other vertices are obtained by numerical minimization. 
We used the \textit{INRIA}'s \textbf{M1QN3} \cite{M1QN3} optimization routine. 
The routine implements a limited memory quasi-Newton 
technique (limited memory Broyden-Fletcher-Goldfarb-Shannon method) of Nocedal \cite{nocedal1980}).



\subsection{Adaptive mesh refinement}

The systems are characterised by two widely different length scales. One is
given by the nematic correlation length and the other by the radius of colloidal particles.
Therefore, we resort to adaptive re-meshing technique in order to obtain
a sufficiently good approximation for the  ``exact''  solution in a reasonable
amount of time and with the finite computing resources at hand. The main objective
of the adaptive refinement is to construct a mesh which meets the given interpolation error 
tolerance for a minimal number of elements. 
Such meshes are called optimally-efficient \cite{dazevedo:1991}.
Let $f$ be one of the components of the  exact solution
$Q_{ij}$ to our problem, and $f^1$  its linear interpolation
on some mesh. It can be shown that the interpolation error associated
with some element $E_k$ (either a tetrahedron or a triangle) satisfies the 
following inequality \cite{george} 
\begin{equation}
 \max_{{\bf x}\in E_k} |f({\bf x}) - f^1({\bf x})| \leqslant  \max_{{\bf x}\in E_k} |x_{\alpha}  {\cal  H} _{\alpha\beta}({\bf x}_0) x_{\beta}|,
\label{error_estimate}
\end{equation}
where ${\bf x}_0$ is some vertex of the element $E_k$,  ${\bf x}$ is the position within element  $E_k$ measured relative to ${\bf x}_0$, and ${\cal H} _{\alpha\beta}({\bf x}_0)  \equiv 
       \frac {\partial ^2 f}
               {\partial x_{\alpha}\partial x_{\beta}}\vert_{{\bf x}_0}$
is the Hessian of the exact solution $f$ evaluated at the vertex ${\bf x_0}$. 
If we define the absolute value of the $2D$ Hessian $|{\cal H}|$ as 
$$
|{\cal H}|= {\cal O}^T\left(\begin{array}{cc}
|\lambda_1|&0\\
0&|\lambda_2| \\
\end{array}\right){\cal O},
$$ 
where ${\cal O}$ is an orthogonal matrix that diagonalizes ${\cal H}$, then
the estimate for the interpolation error on the element $E_k$ can be written in a simpler form
$$ \max_{{\bf x}\in E_k} |f({\bf x}) - f^1({\bf x})| \leqslant  h_k^2 \max (\lambda_{1},\lambda_{2}). $$ 
$h_k$ is the diameter (the length of the longest edge) of the element $E_k$.
 It is straightforward to write down similar expressions for the $3D$ case. 
Therefore, if the sizes of the elements are chosen such that 
$h_k^2 \max (\lambda_{1},\lambda_{2}) \simeq const$ for all the elements, then
the resulting interpolation error will be approximately equally distributed among all the
elements.

As it was mentioned above the {\it BL2D} package supports anisotropic meshes.
The triangulation is governed by specifying a  metric map ${\cal M}$,
and the resulting mesh is characterised by edges of unit length according to the
metric ${\cal M}$ \cite{george}.  We assume that the required metric is proportional to the Hessian, 
${\cal M}\propto|{\cal H}|$. In order to estimate  ${\cal H}$ at some 
vertex ${\bf x}_0)$ we use the  weak definition of the Hessian  \cite{george}
$$
{\cal H}_{\alpha\beta}({\bf x}_0) = \frac{
                                  -\int 
                              \frac{\partial f}{\partial x_\alpha}
                              \frac{\partial {\mathrm \phi}^0}{\partial x_\beta}
                                     {\mathrm d^dr}
                                }
                              {
                              \int {\mathrm \phi}^0  {\mathrm d^dr}
                               }
$$
where the integral is over the elements which share the vertex ${\bf x}_0$, and 
$\phi^0$ is the piecewise linear hat-function associated with
vertex $0$: $\phi^0=1$ at ${\bf x}_0$ and $\phi^0=0$ at any other vertex. We use linear interpolation in order
to obtaine ${\cal H}$ at some internal point ${\bf x}$ of  an element. 

TetGen implements an isotropic mesh refinement strategy which is based
on a maximum local element volume constraint. The constraints on the volume of elements
are obtained by applying the equidistributing principle, where the nodes of the mesh are chosen 
such that for each element $E_k$ the following condition holds: 
\begin{equation}
\int_{E_k} \sqrt{|\det {\cal H}|} d^dr = const.
\label{equidistr_princ}
\end{equation}
In Ref. \cite{dazevedo:1991}, it is shown that an optimally-efficient $2D$ triangulation which minimizes the interpolation error fulfills asymptotically the equidistributing principle (\ref{equidistr_princ}). We assume that the same is valid in the case of $3D$ triangulation.

\footnotesize{
\bibliography{Qdint} 
}

\end{document}